\documentclass[twocolumn,british,english,superscriptaddress,nofootinbib]{revtex4-1}
\usepackage[T1]{fontenc}
\usepackage[latin9]{inputenc}
\usepackage{babel}
\usepackage{array}
\usepackage{units}
\usepackage{multirow}
\usepackage{amsmath}
\usepackage{amssymb}
\usepackage{graphicx}
\usepackage{esint}
\usepackage[unicode=true,pdfusetitle,
 bookmarks=true,bookmarksnumbered=false,bookmarksopen=false,
 breaklinks=false,pdfborder={0 0 1},backref=false,colorlinks=false]
 {hyperref}
\usepackage{breakurl}

\makeatletter

\providecommand{\tabularnewline}{\\}

\RequirePackage{colortbl, tabularx}
\@ifundefined{comment}{}% do nothing if the comment environment is not defined
  {% redefine the comment environment if it is defined
   
  }%

\makeatother

\usepackage{babel}

\makeatother

\begin{document}

\title{Ultrafast destruction and recovery of the spin density wave order
in iron based pnictides: a multi-pulse optical study}

\author{M. Naseska}

\affiliation{Complex Matter Dept., Jozef Stefan Institute, Jamova 39, Ljubljana,
SI-1000, Ljubljana, Slovenia}

\author{A. Pogrebna}

\affiliation{Complex Matter Dept., Jozef Stefan Institute, Jamova 39, Ljubljana,
SI-1000, Ljubljana, Slovenia}

\affiliation{Radboud University, Institute for Molecules and Materials, Nijmegen
6525 AJ, The Netherlands}

\author{G. Cao}

\affiliation{Department of Physics, Zhejiang University, Hangzhou 310027, People\textquoteright s
Republic of China}

\author{Z. A. Xu}

\affiliation{Department of Physics, Zhejiang University, Hangzhou 310027, People\textquoteright s
Republic of China}

\author{D. Mihailovic}

\affiliation{Complex Matter Dept., Jozef Stefan Institute, Jamova 39, Ljubljana,
SI-1000, Ljubljana, Slovenia}

\affiliation{CENN Nanocenter, Jamova 39, Ljubljana SI-1000, Slovenia}

\author{T. Mertelj}

\email{tomaz.mertelj@ijs.si}

\selectlanguage{english}%

\affiliation{Complex Matter Dept., Jozef Stefan Institute, Jamova 39, Ljubljana,
SI-1000, Ljubljana, Slovenia}

\affiliation{CENN Nanocenter, Jamova 39, Ljubljana SI-1000, Slovenia}

\date{\today}

\begin{abstract}
We report on systematic excitation-density dependent all-optical femtosecond
time resolved study of the spin-density wave state in iron-based superconductors.
The destruction and recovery dynamics are measured by means of the
standard and a multi-pulse pump-probe technique. The experimental
data are analyzed and interpreted in the framework of an extended
three temperature model. The analysis suggests that the optical-phonons
energy-relaxation plays an important role in the recovery of almost
exclusively electronically driven spin density wave order.
\end{abstract}
\maketitle

\section{Introduction}

The collectively ordered electronic states are interesting subjects
for driving out of equilibrium by femtosecond optical pulses in order
to get better insight into their nature\cite{iwaiYamamoto2007photoinduced,kublerEhrke2007coherent,schmittKirchmann2008transient,tomeljakSchaefer2009dynamics,petersenKaiser2011,schmittKirchmann2011ultrafast,beaudCaviezel2014time,smallwoodZhang2014time,Madan2016}
and possibly reveal new meta-stable states\cite{ichikawaNozawa2011transient,stojchevskaVaskivskyi2014ultrafast,morrisonChatelain2014photoinduced}
that are not easily reachable by the quasi-equilibrium route. Among
such states is also the orthorhombic antiferromagnetic spin-density-wave
like state in the parent iron-based superconductors compounds\cite{KamiharaKamihara2006,kamiharaWatanabe2008,Stewart2011},
which is interesting not only due to the proximity to the superconducting
state but also due to its collective itinerant nature and relation
to the nematic\cite{ChuangAllan2010,ChuAnalytis2010,DuszaLucarelli2011}
instability.

The ultrafast dynamics of the spin density wave (SDW) state in pnictides
has been extensively studied by various time resolved techniques\cite{MerteljKusar2010,StojchevskaKusar2010,RettigCortes2012,KimPashkin2012,PogrebnaVujicic2014,PatzLi2014,gerberKim2015direct,RettigMariager2016}.
All-optical\cite{PogrebnaVujicic2014} and time resolved (TR) ARPES\cite{RettigCortes2012}
studies show sub-picosecond dynamics with slight slowing down near
the transition temperature, while the orthorhombic lattice splitting
responds much slower\cite{gerberKim2015direct,RettigMariager2016}
upon the ultrafast perturbation. An interesting question is what sets
the sub-picosecond timescale of the suppression and recovery of the
electronic SDW order? At weak suppression it appears that the timescale
is set by the bottleneck in the relaxation of the nonequilibrium electron
distribution function (NEDF) due to the charge gap associated with
the SDW order.\cite{StojchevskaKusar2010,PogrebnaVujicic2014} At
strong suppression the charge-gap bottleneck is suppressed an the
collective SDW dynamics and/or the electron phonon coupling might
play a role in setting the timescale.

In order to improve understanding of the suppression and recovery
timescales at strong suppression we conducted a systematic fluence-dependent
femtosecond time-resolved all-optical study of the SDW state in two
iron-based superconductors parent compounds: AFe$_{2}$As$_{2}$ (A=Eu,Sr).
In the study we supplemented the standard pump-probe technique with
the multipulse technique that proved to be instrumental\cite{yusupovMertelj2010coherent,Madan2016}
to obtain insights into the collective dynamics in charge density
wave systems\cite{yusupovMertelj2010coherent} and superconductors\cite{Madan2016}.

To identify the processes that set the SDW recovery time we analyze
the multipulse data in the framework of an extended three temperature
model (3TM). Surprisingly, the 3TM analysis suggests that an excitation
density dependent \emph{optical-phonons - lattice-bath} energy-relaxation
bottleneck plays a crucial role in the the NEDF relaxation and the
SDW order recovery while the collective SDW order dynamics is too
fast to influence the dynamics beyond $\sim200$ fs. Moreover, the
resilience of the SDW state to strong ultrafast optical excitation
is suggested to be a consequence of a fast electron - optical-phonon
energy transfer during the initial NEDF thermalization on a few hundred
femtosecond timescale that is enhanced at high excitation densities.

\section{Experimental}

\subsection{Samples}

Single crystals of EuFe$_{2}$As$_{2}$ (Eu-122) and SrFe$_{2}$As$_{2}$
(Sr-122) were grown at Zhejiang University by a flux method as described
previously.\cite{PogrebnaVujicic2014} In both compounds the onset
of the antiferromagnetic SDW-like ordering is concurrent with the
structural transition from tetragonal to orthorhombic symmetry $T_{\mathrm{N}}=190$
K for Eu-122\cite{tegelRotter2008} and $T_{\mathrm{N}}=203$ K for
Sr-122.\cite{tegelRotter2008} 

\begin{figure}
\includegraphics[clip,width=0.8\columnwidth]{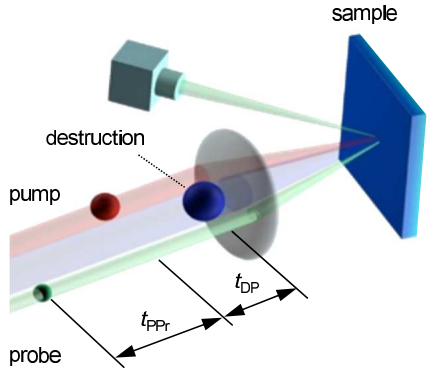}\caption{(Color online) Schematic representation of the three-pulse experiment
and notation of the delays between pulses.}

\label{fig:figShematics} 
\end{figure}

\subsection{Optical setup}

Measurements of the multi-pulse transient reflectivity were performed
using an extension of the standard pump-probe technique, with $\sim50$
fs optical pulses from either 1-kHz or 250-kHz Ti:Al$_{2}$O$_{3}$
regenerative amplifiers seeded with Ti:Al$_{2}$O$_{3}$ oscillators.
The output pulse train was split into destruction (D), pump (P) and
probe (Pr) pulse trains that were independently delayed with respect
to each other. The P and D pulse beams were either at the laser fundamental
($\hbar\omega_{\mathrm{P}}=1.55$ eV) or the doubled ($\hbar\omega_{\mathrm{P}}=3.1$
eV) photon energy, while the Pr beam was always at the the laser fundamental
$\hbar\omega_{\mathrm{pr}}=1.55$ eV photon energy. 

The resulting beams were focused and overlapped on the sample (see
Fig. \ref{fig:figShematics}). As in the standard pump-probe stroboscopic
experiments the multipulse transient reflectivity $\Delta R_{3}/R$
was measured by monitoring the intensity of the weakest Pr beam. The
direct contribution of the unchopped D beam to the total transient
reflectivity, $\Delta R$, was rejected by means of a lock-in synchronized
to the chopper that modulated the intensity of the P beam only. The
fluences of the P and Pr pulses, $\mathcal{\mathcal{F_{\mathrm{Pr}}}<F_{\mathrm{P}}}\lesssim100$~$\mu$J/cm$^{2}$,
were kept in the linear response region. 

\begin{figure*}[tbh]
\includegraphics[clip,width=1\columnwidth]{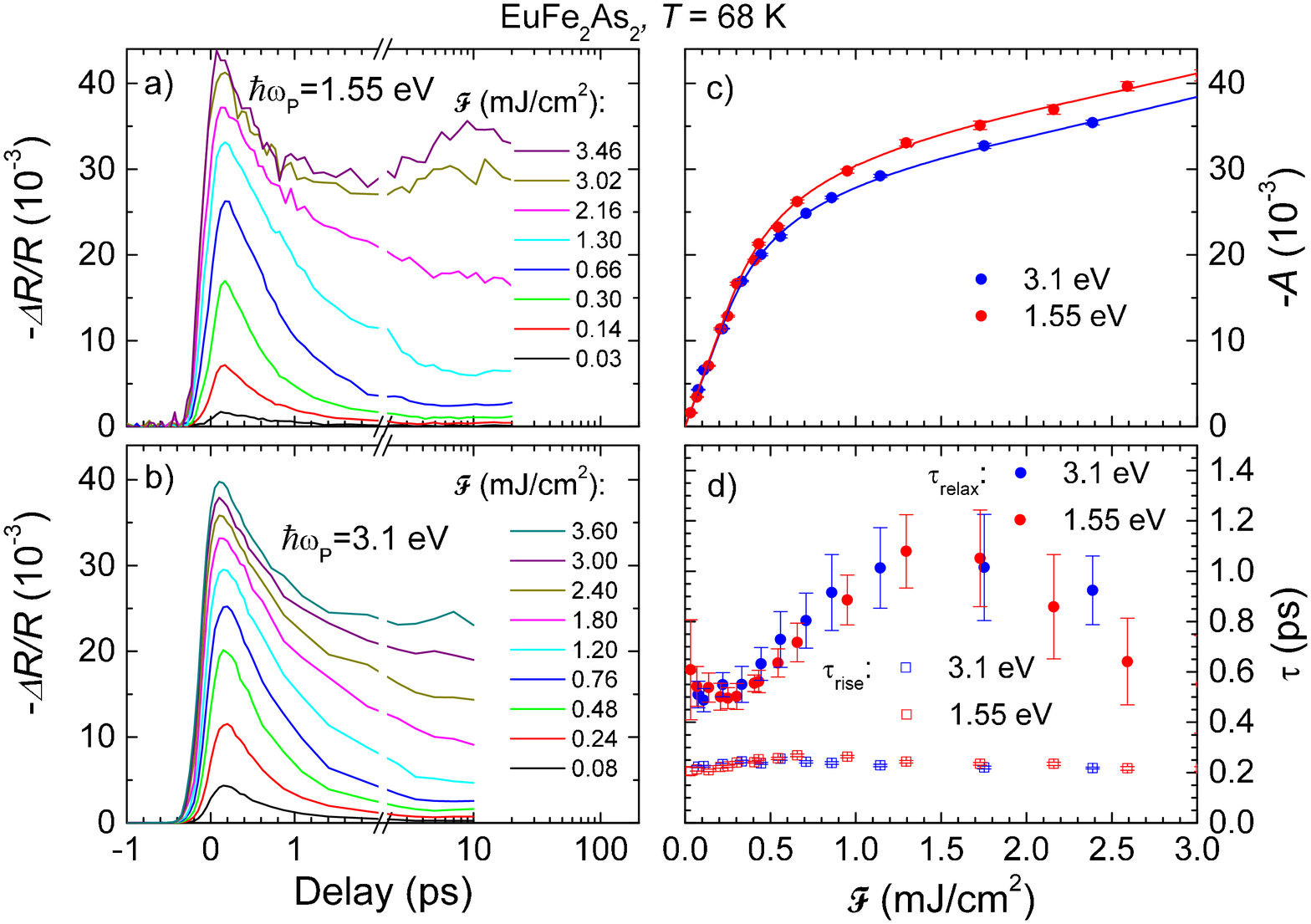}\quad{}\includegraphics[clip,width=1\columnwidth]{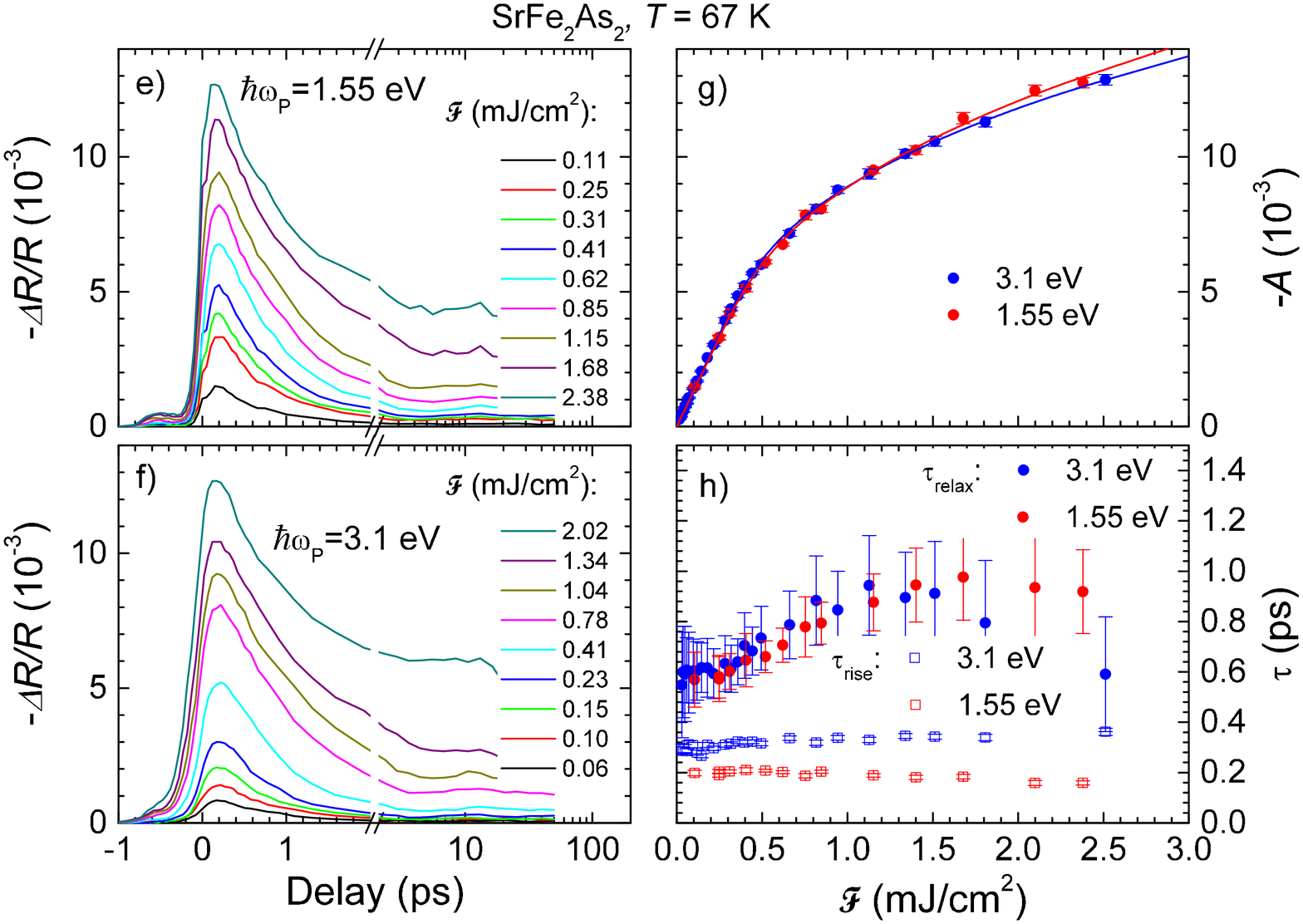}

\caption{(Color online) Fluence dependence of the transient reflectivity in
the SDW state for the $\mathcal{P}^{-}$ probe polarization in Eu-122
(a), (b) and Sr-122 (e), (f) at two different pump-photon energies.
The amplitude of the transient reflectivity as a function of the pump
fluence in Eu-122 (c) and Sr-122 (g). The corresponding transient
reflectivity decay and rise times as a function of the pump fluence
are shown in (d) and (h). The thin lines in (c) and (g) are the saturation
model fits discussed in text.}

\label{fig:fig-F-dep} 
\end{figure*}

Due to the chopping scheme the measured quantity in the multipulse
experiments is the difference between the transient reflectivity in
the presence of P and D pulses, $\Delta R_{\mathrm{DP}}(t_{\mathrm{Pr}},t_{\mathrm{P}},t_{\mathrm{D}})$,
and the transient reflectivity in the presence of the D pulse only,
$\Delta R_{\mathrm{D}}(t_{\mathrm{Pr}},t_{\mathrm{D}})$:

\begin{equation}
\Delta R_{3}(t_{\mathrm{Pr}},t_{\mathrm{P}},t_{\mathrm{D}})=\Delta R_{\mathrm{DP}}(t_{\mathrm{Pr}},t_{\mathrm{P}},t_{\mathrm{D}})-\Delta R_{\mathrm{D}}(t_{\mathrm{Pr}},t_{\mathrm{D}}),\label{eq:DR3}
\end{equation}
where $t_{\mathrm{Pr}}$, $t_{\mathrm{P}}$ and $t_{\mathrm{D}}$
correspond to the Pr, P and D pulse arrival times, respectively. 

When using the doubled P-photon energy the scattered pump photons
were rejected by long-pass filtering, while an \foreignlanguage{british}{analyzer}
oriented perpendicularly to the P-beam polarization was used for rejection
in the case of the degenerate P- and Pr-photon energies. All beams
were nearly perpendicular to the cleaved sample surface (001). Both,
the P and D beam had polarizations perpendicular to the polarization
of the Pr beam, which was oriented with respect to the crystals to
obtain the maximum or minimum amplitude of the sub-picosecond $\Delta R/R$
at low temperatures. The pump beam diameters were, depending on experimental
conditions, in a 50-100 $\mu$m range with somewhat smaller probe
beam diameters. The beam diameters were determined either by a direct
measurement of the profile at the sample position by means of a CMOS
camera or by measuring the transmission through a set of calibrated
pinholes.

\section{Standard pump-probe results}

As noted previously\cite{PogrebnaMertelj2016} we observe a 2-fold
rotational anisotropy of the transient reflectivity with respect to
the probe polarization with different orientation in different domains.
To measure a single domain dominated response the positions on the
sample surface with maximal anisotropy of the response have been chosen
for measurements. In the absence of information about the in-plane
crystal axes orientation in the chosen domains we denote the probe-polarization
orientation according to the polarity of the observed sub-picosecond
low-$T$ response as $\mathcal{P}^{+}$ and $\mathcal{P}^{-}$. The
magnitude of the $\mathcal{P}^{-}$ response is larger than the magnitude
of the $\mathcal{P}^{+}$ response in both compounds so in the multi-pulse
experiments the $\mathcal{P}^{-}$ Pr polarization was used in most
of the cases.

\subsection{Fluence dependence}

In Fig. \ref{fig:fig-F-dep} we plot the fluence dependence of the
standard 2-pulse transient reflectivity in the SDW state. In both
compounds we observe a linear scaling of $\Delta R/R$ with the pump
fluence ($\mathcal{F}_{\mathrm{P}}$) up to the threshold fluence,
$\mathcal{F}_{\mathrm{th}}\sim0.2$ mJ/cm$^{2}$. Above this value
the amplitude of the initial sub-ps transient shows a partial saturation
increasing linearly with a different slope and nonzero intercept above
$\mathcal{F}_{\mathrm{P}}\sim1$ mJ/cm$^{2}$. In this region of fluence
also a long lived component following the initial sub-ps transient
becomes rather prominent. 

The risetime of the transients, $\tau_{\mathrm{rise}}$, shows no
fluence dependence while the initial sub-ps decay time rises from
the below-$\mathcal{F}_{\mathrm{th}}$-value of $\tau_{\mathrm{relax}}\sim0.6$
ps to a maximum of $\tau_{\mathrm{relax}}\sim1$ ps at $\mathcal{F}_{\mathrm{P}}\sim1.5$
mJ/cm$^{2}$ decreasing back to $\tau_{\mathrm{relax}}\sim0.6$ ps
at the highest $\mathcal{F}_{\mathrm{P}}\sim3$ mJ/cm$^{2}$.

\subsection{Transient heating}

\begin{figure}
\includegraphics[clip,width=0.75\columnwidth]{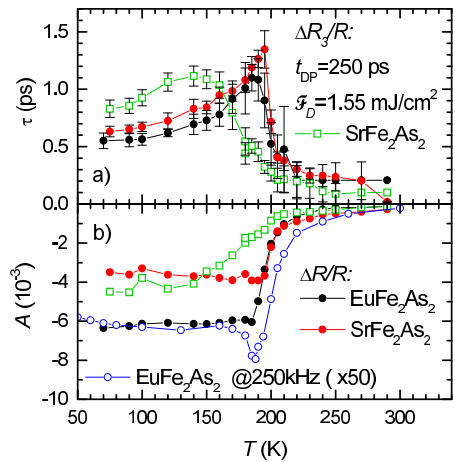}\caption{(Color online) Temperature dependence of the relaxation time and amplitude
for the $\mathcal{P}^{-}$ polarization measured at 1-kHz pulse repetition
rate in absence (red and black full circles) and presence of the D-pulse
heating (green open squares). For comparison we plot also the amplitude
$T$-dependence measured at 250-kHz repetition rate (blue open circles)
at $\sim50$ times lower pump fluence.}

\label{fig:fig-A-vs-T} 
\end{figure}

In order to experimentally assess the transient thermal heating of
the experimental volume we measured temperature dependence of $\Delta R_{3}/R$
in Sr-122 at $\mathcal{F}_{\mathrm{D}}=1.55$ mJ/cm$^{2}$ and long
$t_{\mathrm{DP}}\sim$250 ps,\footnote{The electronic system and the lattice are expected to be in the local
thermal equilibrium on this time scale.} and compared it to temperature dependence of $\Delta R/R$ in the
absence of the D pulse. From Fig. \ref{fig:fig-A-vs-T} we can see
that in the absence of the D pulse the relaxation time shows a characteristic\cite{PogrebnaVujicic2014}
$T$-dependence and can be used as a proxy to the temperature to estimate
the transient lattice heating in the presence of the D pulse. In the
presence of the D-pulse the characteristic relaxation-time peak at
$T_{\mathrm{N}}$ is shifted $\sim60$ K towards lower temperature
and smeared due to the temperature gradient perpendicular to the sample
surface. The experimental thermal heating at $T\sim150$ K is therefore
$\Delta T\sim60$ K at $\mathcal{F}_{\mathrm{D}}=1.55$ mJ/cm$^{2}$
increasing to $\Delta T\sim90$ K at $T\sim70$~K. 

On the other hand, taking into account the experimental temperature-dependent
specific heat capacity\cite{Herero-MartinScagnoli2009,ChenLi2008}
and optical\cite{WuBarisic2009,CharnukhaLarkin2013} data we estimate\footnote{We obtain the light penetration depth of $\alpha_{\mathrm{Pr}}^{-1}=27$
nm and $24$ nm in Eu-122 and $24$ nm and 17 nm in Sr-122 at $\hbar\omega=1.55$
eV and $\hbar\omega=3.1$ eV , respectively.} (at $T\sim70$~K) a temperature increase of $\mbox{\ensuremath{\Delta}}T\sim210$
K at the fluence $\mbox{\ensuremath{\mathcal{F_{\mathrm{D}}}}}=1.55$~mJ/cm$^{2}$,
while the transition temperature of $T_{\mathrm{N}}\sim200$ K would
be reached at $\mbox{\ensuremath{\mathcal{F_{\mathrm{P}}}}}\sim1$~mJ/cm$^{2}$.
The estimated $\mbox{\ensuremath{\Delta}}T$ is therefore more than
two times larger than the directly measured.

\begin{figure}[tbh]
\includegraphics[clip,width=0.7\columnwidth]{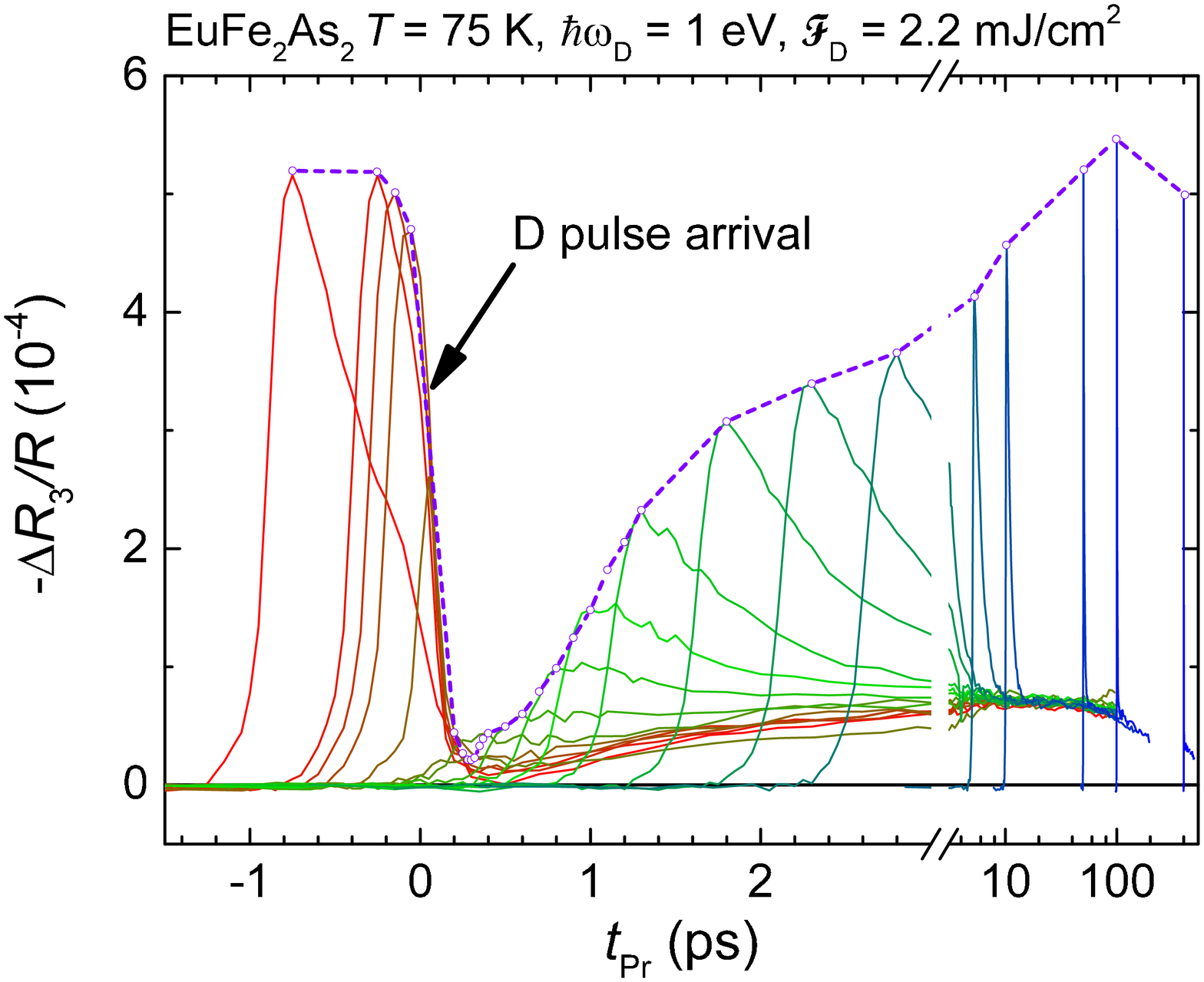}\caption{(Color online) The $\mathcal{P}^{-}$ multi-pulse transient reflectivity
in Eu-122 in the presence of the destruction pulse arriving at $t_{\mathrm{D}}=0$
ps for different pump pulse arrival times measured with the 250-kHz
repetition rate laser system. The dashed line represents the trajectory
defined in text.}

\label{fig:fig-trajectory} 
\end{figure}

\begin{figure}
\includegraphics[width=0.7\columnwidth]{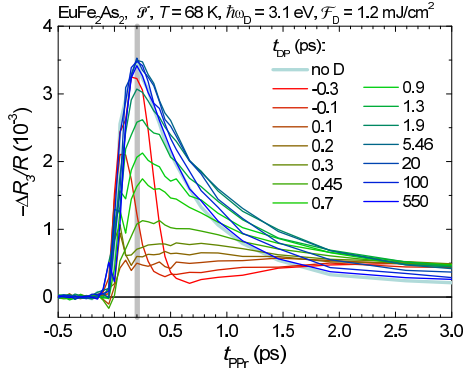}

\caption{(Color online) Example of the multi-pulse transient reflectivity in
Eu-122 at different $t_{\mathrm{DP}}$ plot relative to the pump arrival
time, $t_{\mathrm{P}}$, measured with the 1-kHz repetition rate laser
system. The vertical gray line indicates the trajectory readout pump-probe
delay.\label{fig:readout}}
\end{figure}

\section{Multi pulse results}

\subsection{Multi-pulse trajectories}

In Fig. \ref{fig:fig-trajectory} we plot results of a typical multi-pulse
experiment where the destruction pulse arrives at $t_{\mathrm{D}}=0$
ps, while the pump-pulse and probe-pulse arrival times are varied.
By tracking the value of $\Delta R_{3}(t_{\mathrm{Pr}},t_{\mathrm{P}},t_{\mathrm{D}})$
at a constant $t_{\mathrm{PPr}}=t_{\mathrm{Pr}}-t_{\mathrm{P}}$ at
the extremum ($t_{\mathrm{PPr}}\sim200$ fs) of the unperturbed\footnote{In the absence of the D pulse.}
$\Delta R/R$ (Fig. \ref{fig:readout}), we define the trajectory,
$A_{3}(t_{\mathrm{DP}})$, where $t_{\mathrm{DP}}=t_{\mathrm{P}}-t_{\mathrm{D}}$
is the delay between the D and P pulse. Due to the finite $t_{\mathrm{PPr}}$
at the readout of $A_{3}(t_{\mathrm{DP}})$ the temporal resolution
of the trajectory is limited to $\sim t_{\mathrm{PPr}}\approx200$
fs.

\begin{figure}
\medskip{}

\includegraphics[clip,height=0.85\columnwidth]{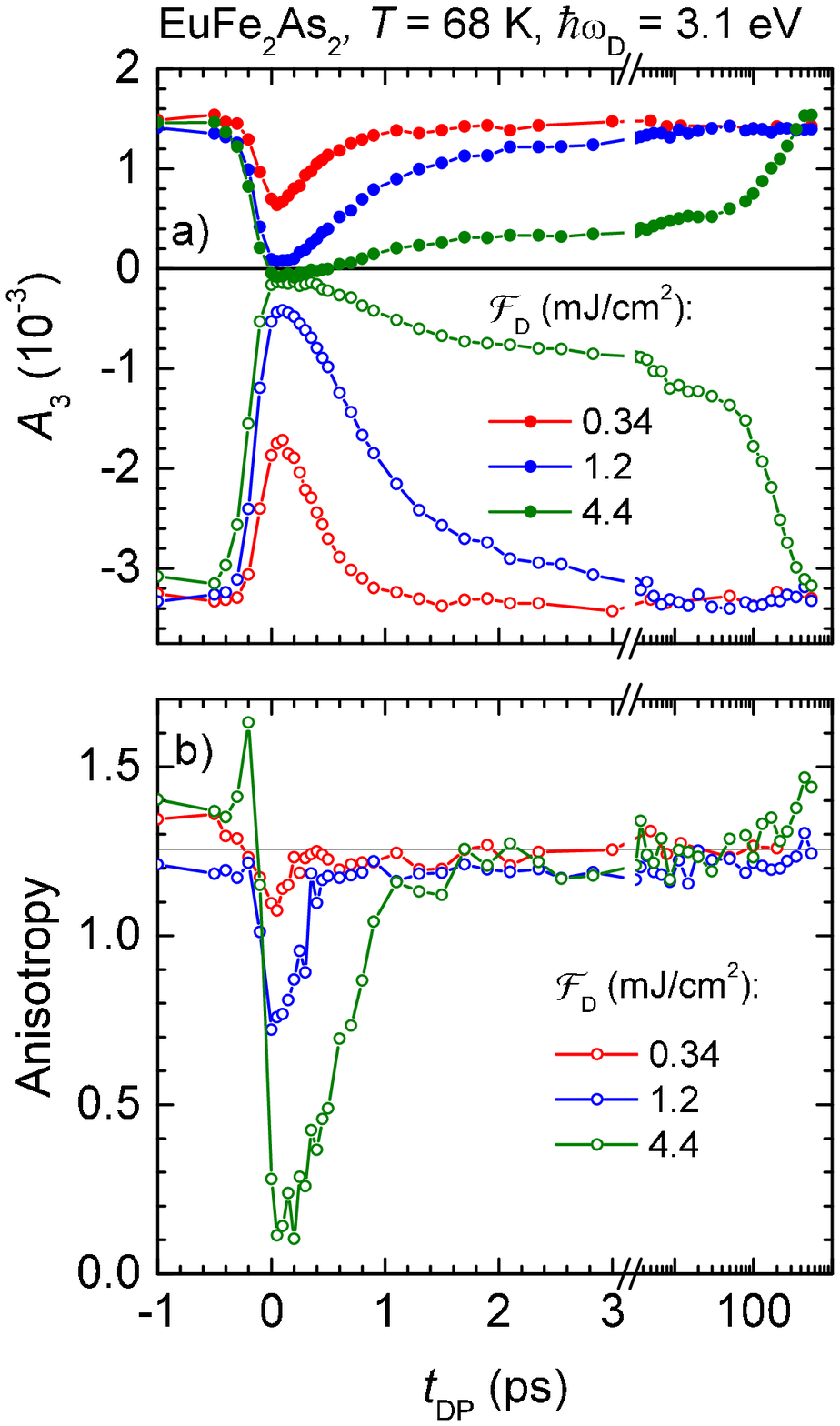} \includegraphics[clip,height=0.85\columnwidth]{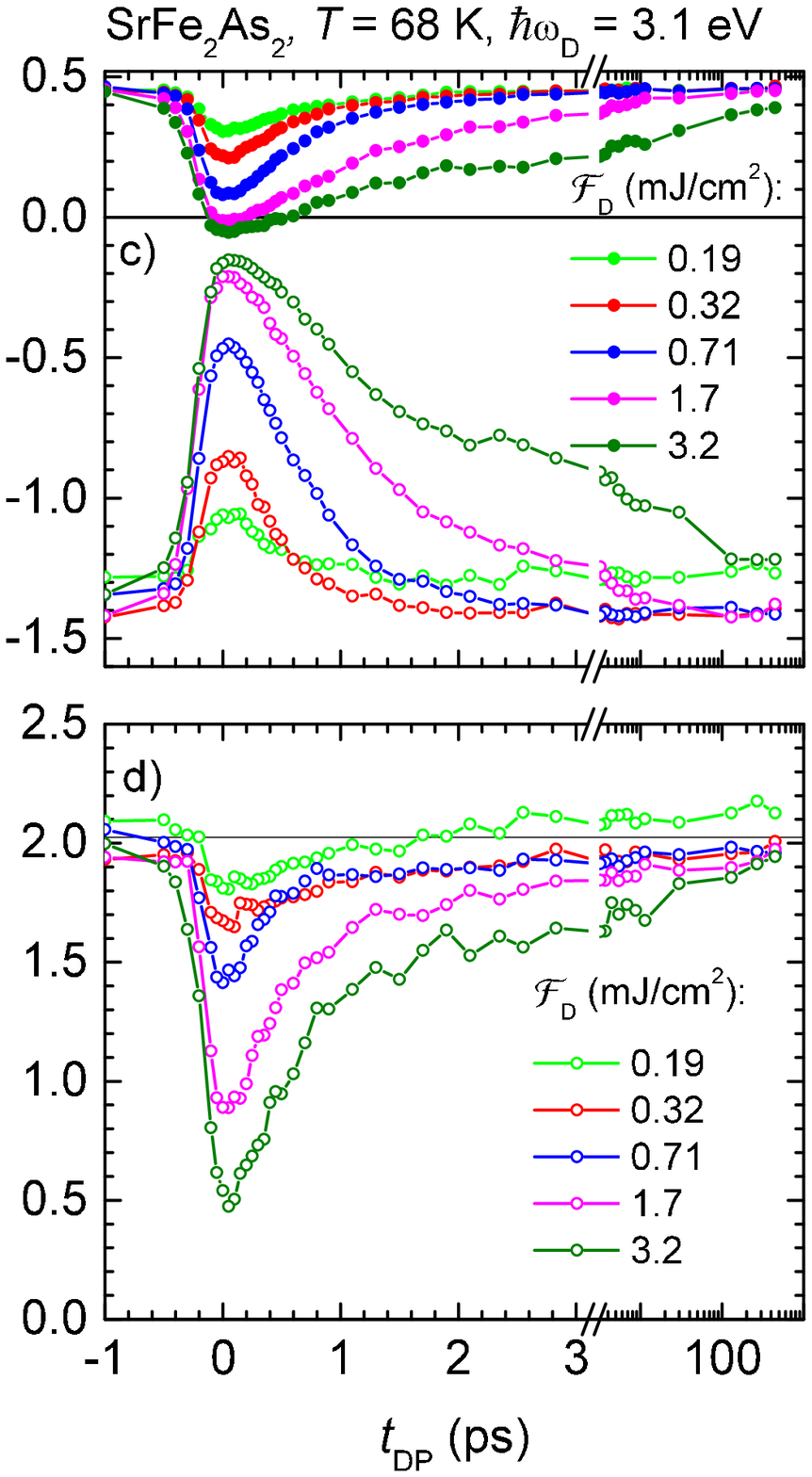}\caption{(Color online) The trajectories for the $\mathcal{P}^{+}$ (full symbols)
and $\mathcal{P}^{-}$ (open symbols) polarizations at a few characteristic
destruction-pulse fluences in Eu-122 (a) and Sr-122 (c). Anisotropy
(see text for definition) of the trajectories at different destruction
fluences in Eu-122 (b) and Sr-122 (d).}

\label{fig:fig-anis} 
\end{figure}

\begin{figure*}[tbh]
\includegraphics[angle=-90,width=1\columnwidth]{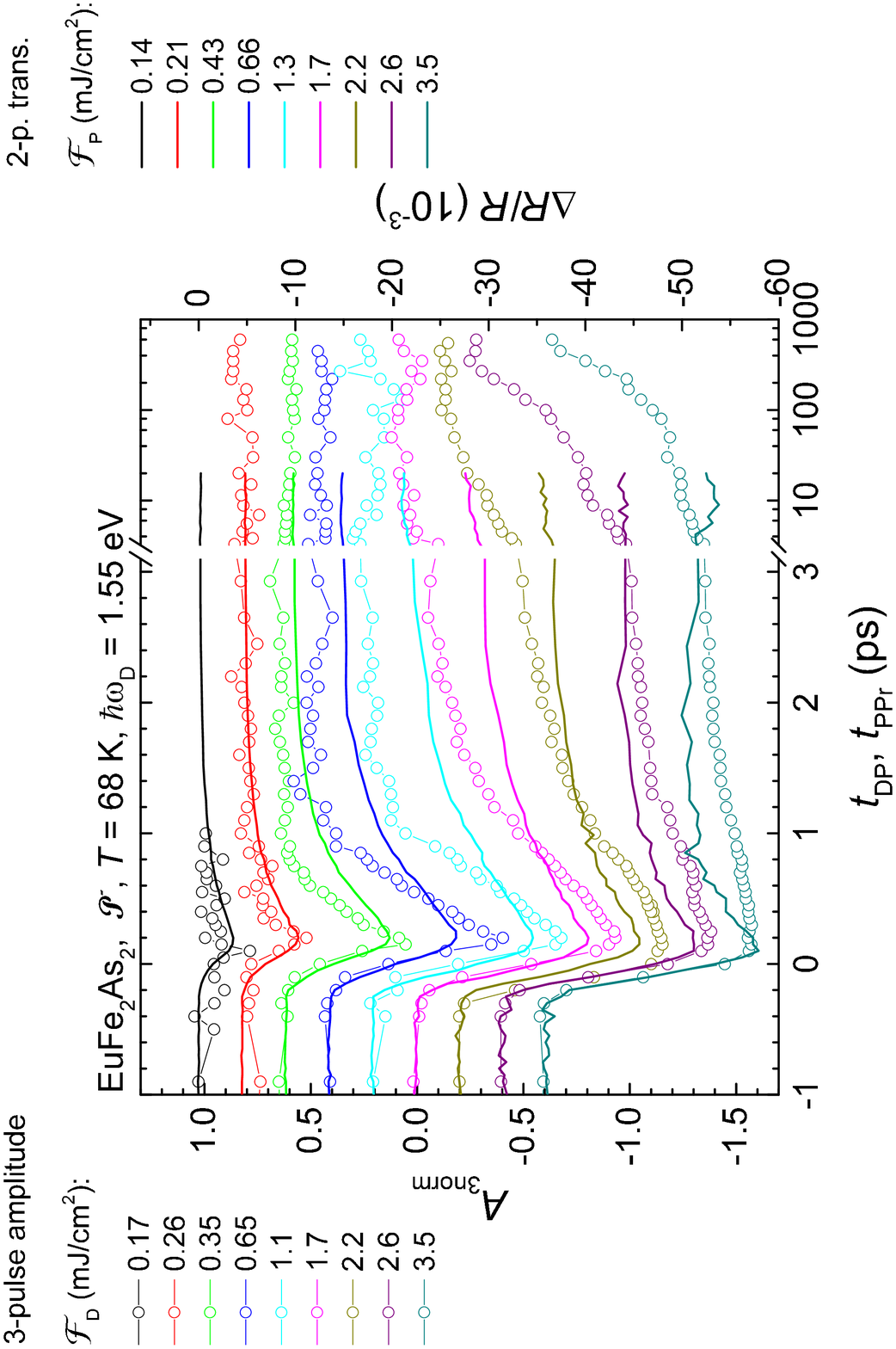}\includegraphics[clip,angle=-90,width=1\columnwidth]{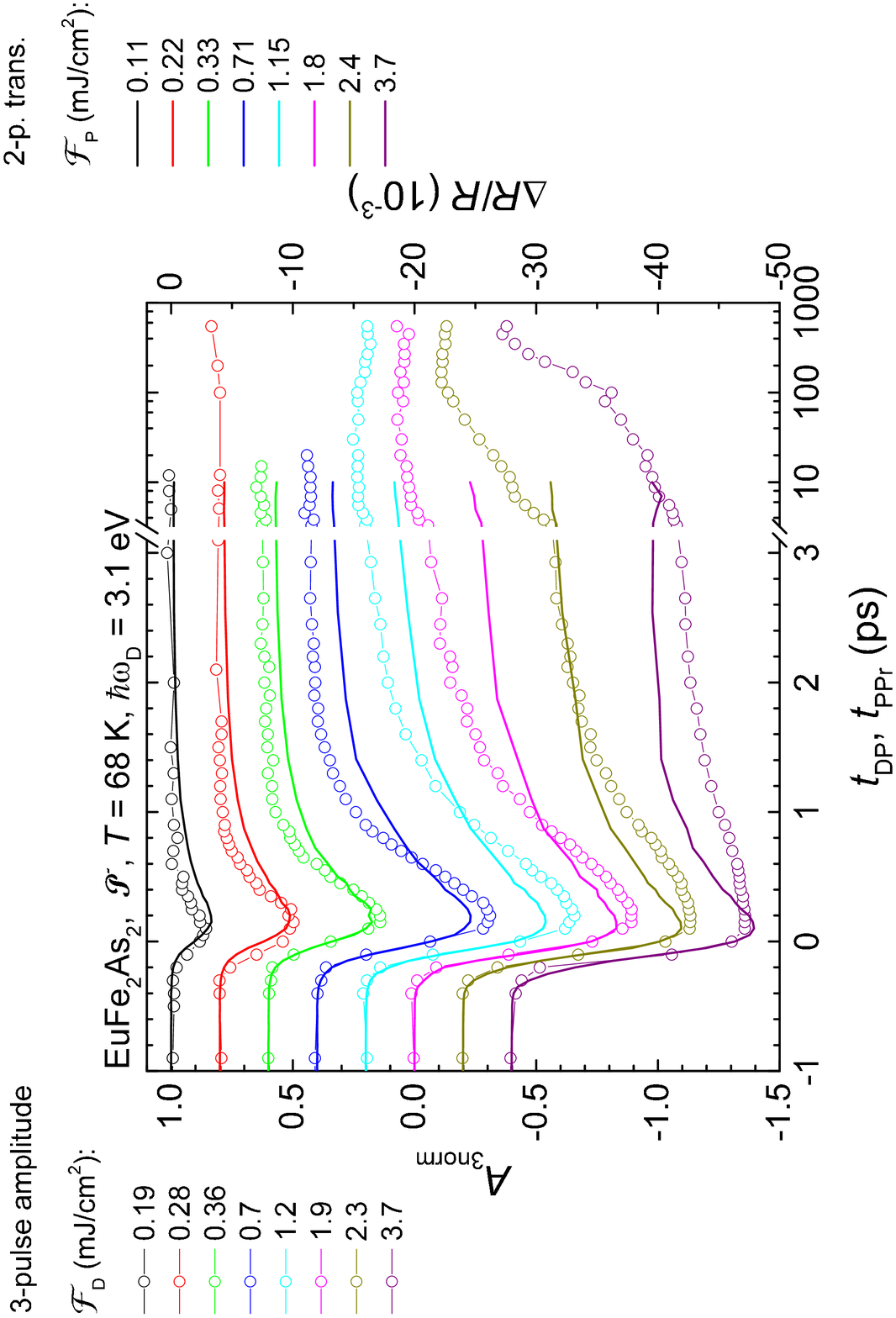}\caption{(Color online) Comparison of the normalized $\mathcal{P}^{-}$ trajectories
(open symbols) to the $\mathcal{P}^{-}$ transient reflectivity (full
lines) at different destruction/pump fluences in Eu-122 for two different
pump/destruction photon energies.}

\label{fig:fig-3p-vs-2p-efa} 
\end{figure*}

In Fig. \ref{fig:fig-anis} (a) and (c) we plot typical trajectories
for both probe polarizations at $T\sim70$~K. Below $\mathcal{F}_{\mathrm{D}}\sim1$
mJ/cm$^{2}$ the trajectories indicate a recovery of the ordered state
on the sub-ps timescale. Above $\mathcal{F}_{\mathrm{D}}\sim2$ mJ/cm$^{2}$
the recovery timescale slows down beyond hundreds of picoseconds.
In the intermediate region 1 mJ/cm$^{2}$ $\lesssim\mathcal{F}_{\mathrm{D}}\lesssim2$
mJ/cm$^{2}$ the recovery is still observed on a few ps timescale.
Since the heat can not diffuse out of the excited sample volume on
this timescale this indicates that the transient lattice temperature
does not exceed $T_{\mathrm{N}}$ below $\mathcal{F}_{\mathrm{D}}\sim2$
mJ/cm$^{2}$. This fluence therefore represents the boundary between
the fast-quench and the slow-quench conditions.

In Fig. \ref{fig:fig-anis} (b) and (d) we plot also the anisotropy
defined as ($A_{3\mathcal{P}+}-A_{3\mathcal{P}-})/(A_{3\mathcal{P}+}-A_{3\mathcal{P}-})$.
In Eu-122 the anisotropy recovers on the sub-ps timescale even at
the slow quench conditions while in Sr-122 the initial sub-ps recovery
is followed by a slower tail lasting more than $\sim10$ ps. 

In Fig. \ref{fig:fig-3p-vs-2p-efa} we also compare the trajectories
to the standard transient reflectivity measured at similar excitation
fluences. In the case of fast quench, $\mathcal{F}_{\mathrm{D}}\lesssim2$
mJ/cm$^{2}$, the trajectories recover faster than the corresponding
transient reflectivity for both D-photon energies. In the case of
extremely slow quench, $\mathcal{F}_{\mathrm{D}}\gtrsim2$ mJ/cm$^{2}$,
the trajectory dynamics shows only the slow recovery while the transient
reflectivity still displays a partial initial sub-picosecond relaxation. 

\begin{figure}[tbh]
\medskip{}

~\includegraphics[clip,width=0.65\columnwidth]{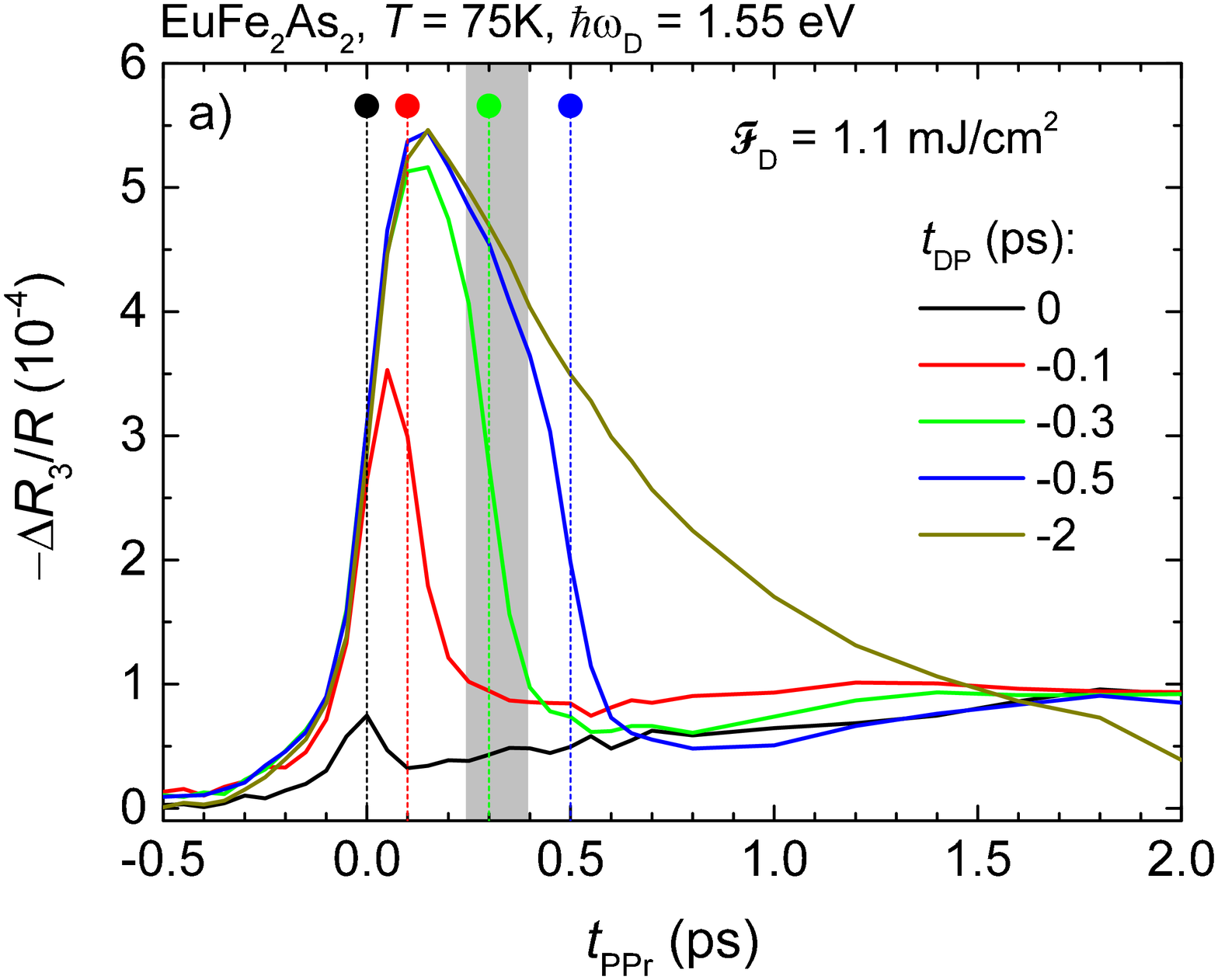}\medskip{}

\includegraphics[width=0.65\columnwidth]{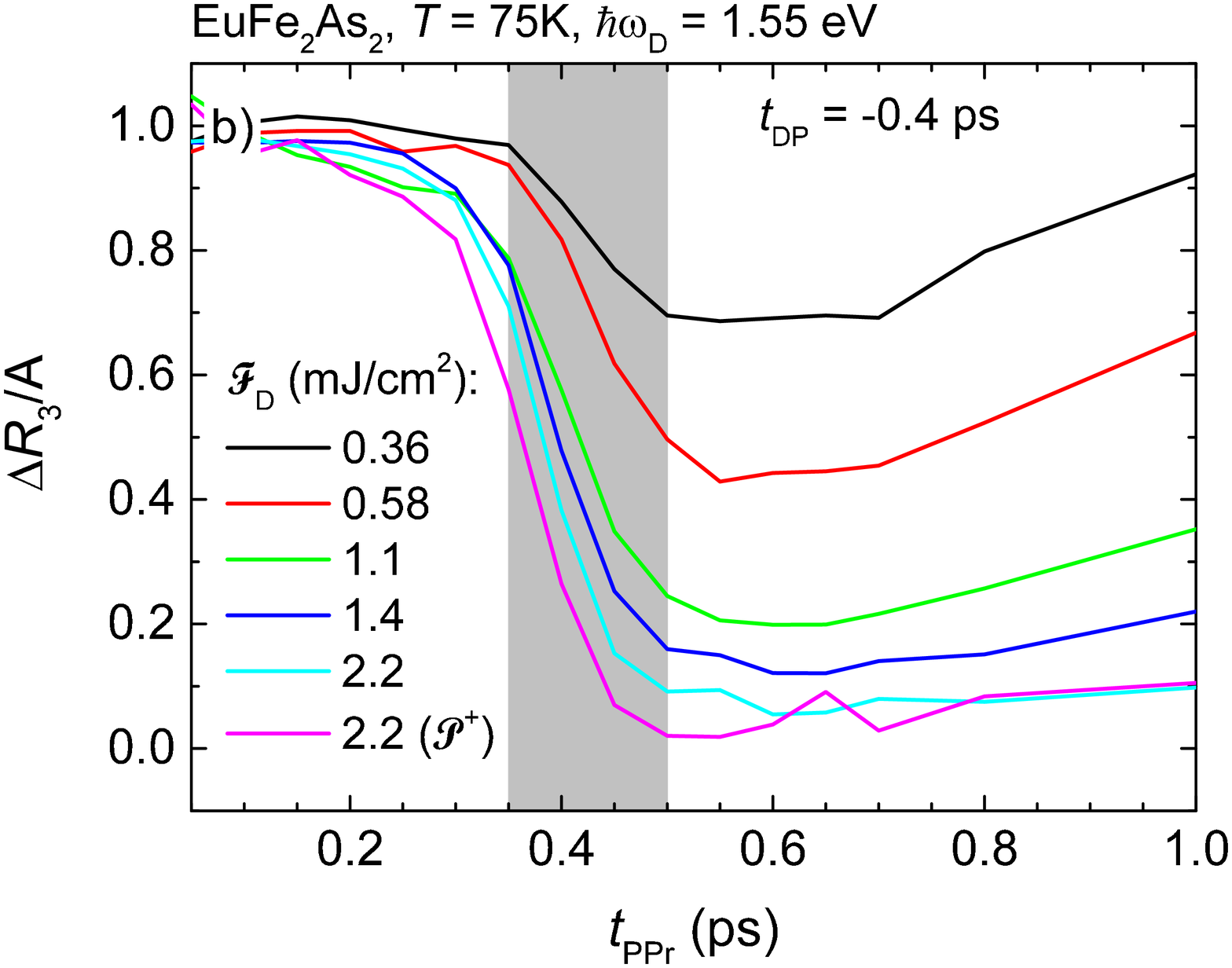}

\caption{(Color online) (a) The $\mathcal{P}^{-}$ multi-pulse transient reflectivity
in Eu-122 at different destruction pulse arrival times $t_{\mathrm{DP}}$.
The timescale of the transient reflectivity suppression region of
150~fs for $t_{\mathrm{DP}}=0.3$~ps is indicated by the shaded
region. (b) The relative suppression of the $\mathcal{P}^{-}$ $\Delta R_{3}$
as a function of $\mathcal{F}_{\mathrm{D}}$ for the destruction pulse
arriving at $t_{\mathrm{DP}}=-0.4$~ps.}

\label{fig:fig-Destruction} 
\end{figure}

\subsection{Destruction timescale}

To determine the destruction timescale of the ordered state we chose
a negative $t_{\mathrm{DP}}$ and analyze the suppression of $\Delta R_{3}/R$
after the D pulse arrival. As shown in Fig. \ref{fig:fig-Destruction}
$\Delta R_{3}/R$ is suppressed within $\sim150$ fs at $1.55$-eV
D-photon energy. The suppression timescale does not depend on $\mathcal{F}_{\mathrm{D}}$
although we observe an earlier onset of the suppression at higher
$\mathcal{F}_{\mathrm{D}}$. The effect can be attributed to the wing
of the D pulse extending beyond $\sim50$ fs that at higher $\mathcal{F}_{\mathrm{D}}$
contains enough energy to start the ordered-state suppression prior
to the arrival of the central part of the D pulse.

Comparing the trajectories measured at different destruction-photon
energies in Fig.~\ref{fig:fig-3p-vs-2p-efa} we observe a sharper
feature around the maximal-suppression $t_{\mathrm{DP}}$ in the case
of the degenerate, $\hbar\omega_{\mathrm{D}}=1.55$ eV, D-photon energy.
While the suppression timescale appears identical for both D-photon
energies in Eu-122 the suppression at 3.1-eV D-photon energy in Sr-122
appears slower {[}Fig. \ref{fig:fig-3T-sim-400} (b){]} consistent
with the slower risetime in the standard pump-probe experiment {[}Fig.
\ref{fig:fig-F-dep} (h){]}.

\section{Analysis and Discussion}

\subsection{Destruction-pulse absorption saturation}

The large difference between experimentally determined transient lattice
heating and the estimate based on the equilibrium optical and thermodynamic
properties indicates that the D-pulse energy $\sim250$ ps after the
pulse arrival is deposited in a layer that is about $\sim3$ times
thicker than the optical penetration depth ($\sim20-30$ nm) at the
highest fluences used. This large energy deposition depth can neither
be accounted for by the thermal diffusion\footnote{In the absence of thermal transport data we approximate $c$-axis
thermal conductivity using Wiedmann-Franz law taking the measured
inter-plane resistivity and heat capacity data\cite{ChenLi2008} in
SrFe$_{2}$As$_{2}$ to obtain thermal diffusivity of $K_{z}\sim$0.01
nm$^{2}$/ps.} on the $\sim250$ ps timescale nor the initial ballistic photoexcited
carrier transport\footnote{The in-plane mean carrier free path in Sr-122 at low $T$ is of the
order of a few 100 nm with the Fermi velocity of $\sim5\times10^{4}$
m/s.\cite{SutherlandHills2011} While on the 1-ps timescale, when
the electronic system is still highly excited, the in-plane ballistic
transport on the 100 nm length scale would be possible, the out-of-plane
transport on a similar length scale at elevated $T\sim100$ K is impossible
due to the substantial resistivity anisotropy\cite{ChenLi2008} of
the order of $\sim100$ in Sr-122.}. The most plausible explanation that remains is therefore saturation
of absorption. This is supported also by the fact that the multipulse
trajectories (see Fig. \ref{fig:fig-3p-vs-2p-efa}) show a sharp feature
near $t_{\mathrm{DP}}=0$ when the pump and destruction photon energies
are degenerate.

\subsection{Three temperature model simulations of recovery}

In all-optical experiments it is generally not possible to \emph{directly}
disentangle dynamics of different degrees of freedom due to unknown
response functions. In general, both $\Delta R$ and $A_{3}$ can
couple to single particle and order parameter excitations. We therefore
seek better insight into the recovery by means of semi-empirical simulations
similar as previously in the cuprate superconductors.\cite{Madan2016,madanBaranov2017} 

At low excitation densities the order parameter as well $\Delta R$
can usually be linearly expanded in terms of a single parameter\footnote{In the case of the Rothwarf-Taylor bottleneck model the parameter
is the nonequilibrium quasiparticle density.} that is used to describe the nonequilibrium electronic distribution
function (NEDF) dynamics. In the present compounds we have conjectured
that the low-excitation transient reflectivity couples to the collective
SDW order parameter that has fast femtoseconds-timescale dynamics.
In such case the order parameter and transient reflectivity directly
follow the magnon-bottleneck governed NEDF dynamics.\cite{PogrebnaVujicic2014} 

At high excitation densities the relation between the NEDF and order
parameter becomes nonlinear and the simple low-excitation description
of $\Delta R$ is expected to break down. This is indicated by the
difference between the relaxation dynamics (Fig. \ref{fig:fig-3p-vs-2p-efa})
observed in the standard pump-probe and multi-pulse experiments that
suggests that NEDF and the order parameter have different delay dependence. 

\begin{table*}
\begin{tabular}{>{\raggedright}m{3.7cm}|c|c|c|c|c|c|c|c}
 & $\eta$\footnote{Fixed at the selected values for the case of TR-ARPES.} & $\gamma_{\mathrm{e}}$ & $G_{\mathrm{eo}}$  & $\lambda\left\langle \omega^{2}\right\rangle $ & $c_{\mathrm{E}0}$\footnote{$T_{\mathrm{E}}$ was without fitting set to 300K.}  & $G_{\mathrm{ol}}$ & $k=\kappa/V_{\mathrm{mol}}$  & $L_{\mathrm{D}}/\mathcal{F}_{\mathrm{D}}$\footnote{Obtained from the two highest fluences fit in the multipulse case.}\tabularnewline
 & - & mJ/mol K$^{2}$  & TW/mol K  & (meV)$^{2}$ & J/mol K  & TW/mol K  & W/m K & nm cm$^{2}$/mJ\tabularnewline
\hline 
\multirow{2}{3.7cm}{Eu-122 (TR-ARPES)\cite{RettigCortes2013} \\
($T\sim100$ K) \footnote{ $\mathcal{F}_{\mathrm{D}}\sim1$ mJ/cm$^{2}$}} & $0.01$ & $52\pm3$ & $34\pm3$ & $39\pm3$ & $67\pm14$ & $7\pm5$ & - & 17\tabularnewline
\cline{2-9} 
 & 0.5 & $30\pm2$ & $23\pm3$ & $45\pm3$ & $76\pm20$ & $11\pm9$ & - & 17\tabularnewline
\hline 
Eu-122 (present work)\\
($T\sim70$ K)  & Fig. \ref{fig:3T-fit-par} & $49\pm1$ & $13\pm1$ & $16\pm2$ & $10\pm1$ & \multirow{1}{*}{Fig. \ref{fig:3T-fit-par}} & \multirow{1}{*}{$14\pm1$} & \multirow{1}{*}{$17\pm5$}\tabularnewline
\hline 
Sr-122 (present work)\\
($T\sim70$ K)  & Fig. \ref{fig:3T-fit-par} & $57\pm2$ & $16\pm1$ & $17\pm2$ & $24\pm1$ & \multirow{1}{*}{Fig. \ref{fig:3T-fit-par}} & \multirow{1}{*}{$10\pm1$} & \multirow{1}{*}{$23\pm5$}\tabularnewline
\end{tabular}\caption{Comparison of the three temperature model parameters obtained from
fits. The detailed definition of the parameters is given in Section
\ref{sub:Three-temperature-model}.\label{tab:3Tpara}}
\end{table*}

In the cuprate superconductors the characteristic timescale of the
order parameter relaxation appears to be\cite{Madan2016,madanBaranov2017}
on a picosecond timescale and the intrinsic order parameter dynamics
plays an important role on the experimental-observation timescale.
In the present case the SDW order parameter is expected to relax much
faster due to a larger gap ($2\Delta_{\mathrm{SDW}}\sim200$ meV\cite{PogrebnaVujicic2014}).
An estimate of the SDW amplitude mode frequency,\cite{psaltakis1984}
$\omega_{\mathrm{AM}}=2\Delta_{\mathrm{SDW}}/\hbar$, would lead to
the relaxation timescale bottom limit of $\sim3$ fs. On the other
hand, the SDW transition is coupled to the structural transition that
could lead to renormalization and slowdown of the order parameter
relaxation timescale. Recent time resolved X-ray diffraction experiments\cite{gerberKim2015direct,RettigMariager2016}
showed, however, that on the tens of picoseconds timescale the orthorhombic
lattice splitting is decoupled from the electronic order parameter.

In the present multi-pulse experiments the time resolution of the
trajectories is not better than the risetime of the standard pump-probe
response ($\sim200$ fs). Any intrinsic order parameter dynamics faster
than $\sim200$ fs would therefore not be revealed in the experiment.
Since it is very likely that the intrinsic order parameter relaxation
timescale is faster than the resolution we check this hypothesis by
simulating the trajectories assuming that the order parameter and
the optical response directly follow the NEDF on the experimentally
accessible timescales. 

Since modeling of the NEDF dynamics in strongly excited collectively
ordered systems such are SDWs is prohibitively difficult we further
assume\cite{RettigCortes2013} that NEDF can be approximately described
by an electronic temperature. To calculate the optical response we
use an empirical response function assuming and that the amplitude
of the pump-pulse induced transient dielectric constant, $\Delta\epsilon_{3}^{\mathrm{A}}(z,t_{\mathrm{DP}})$,
depends on the local electronic temperature, $T_{\mathrm{e}}(z,t_{\mathrm{DP}})$,
only,

\begin{equation}
\Delta\epsilon_{3}^{\mathrm{A}}(z,t_{\mathrm{DP}})\propto A(T_{\mathrm{e}}[z,t_{\mathrm{DP}}]).\label{eq:response-function}
\end{equation}
Here $A(T)$ is the experimental $T$-dependent amplitude measured
in the absence of the D pulse shown in Fig. \ref{fig:fig-A-vs-T}
(b) and $z$ corresponds to the normal distance from the sample surface.
For the sake of simplification any radial dependence is neglected.
The multipulse transient reflectivity amplitude is given by (see Appendix
\ref{sub:Transient-reflectivity}): \begin{widetext} 
\begin{eqnarray}
A_{3}(t_{\mathrm{DP}}) & \propto\int_{0}^{\infty}dz & e^{-\alpha_{\mathrm{Pr}}z}\cos\left(2n_{\mathrm{Pr}}\frac{\omega_{0}}{c_{0}}z-\phi\right)\Delta\epsilon_{3}^{\mathrm{A}}(z,t_{\mathrm{DP}}),\label{eq:sat-dr-2D}
\end{eqnarray}
\end{widetext}where $\alpha_{\mathrm{Pr}}$ and $n_{\mathrm{Pr}}$
are the probe absorption coefficient and the real part of the refraction
index, respectively. The phase $\phi$ depends (see Appendix, Eq.
(\ref{eq:phi})) on the static complex refraction index and the ratio
between the real and imaginary part of $\Delta\epsilon_{3}^{\mathrm{A}}$.

\begin{figure*}[tbh]
\includegraphics[clip,width=1\columnwidth]{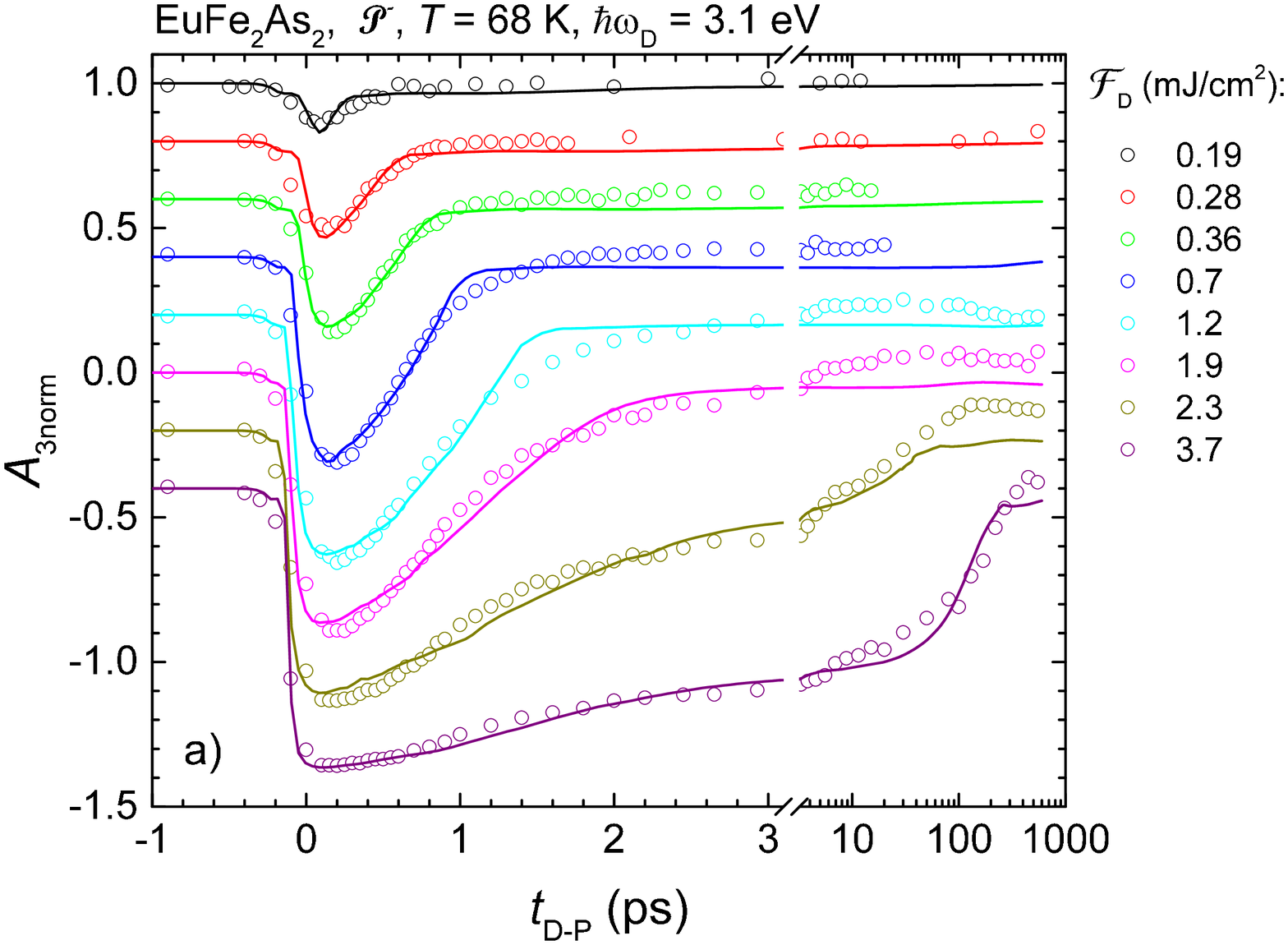}\includegraphics[clip,width=1\columnwidth]{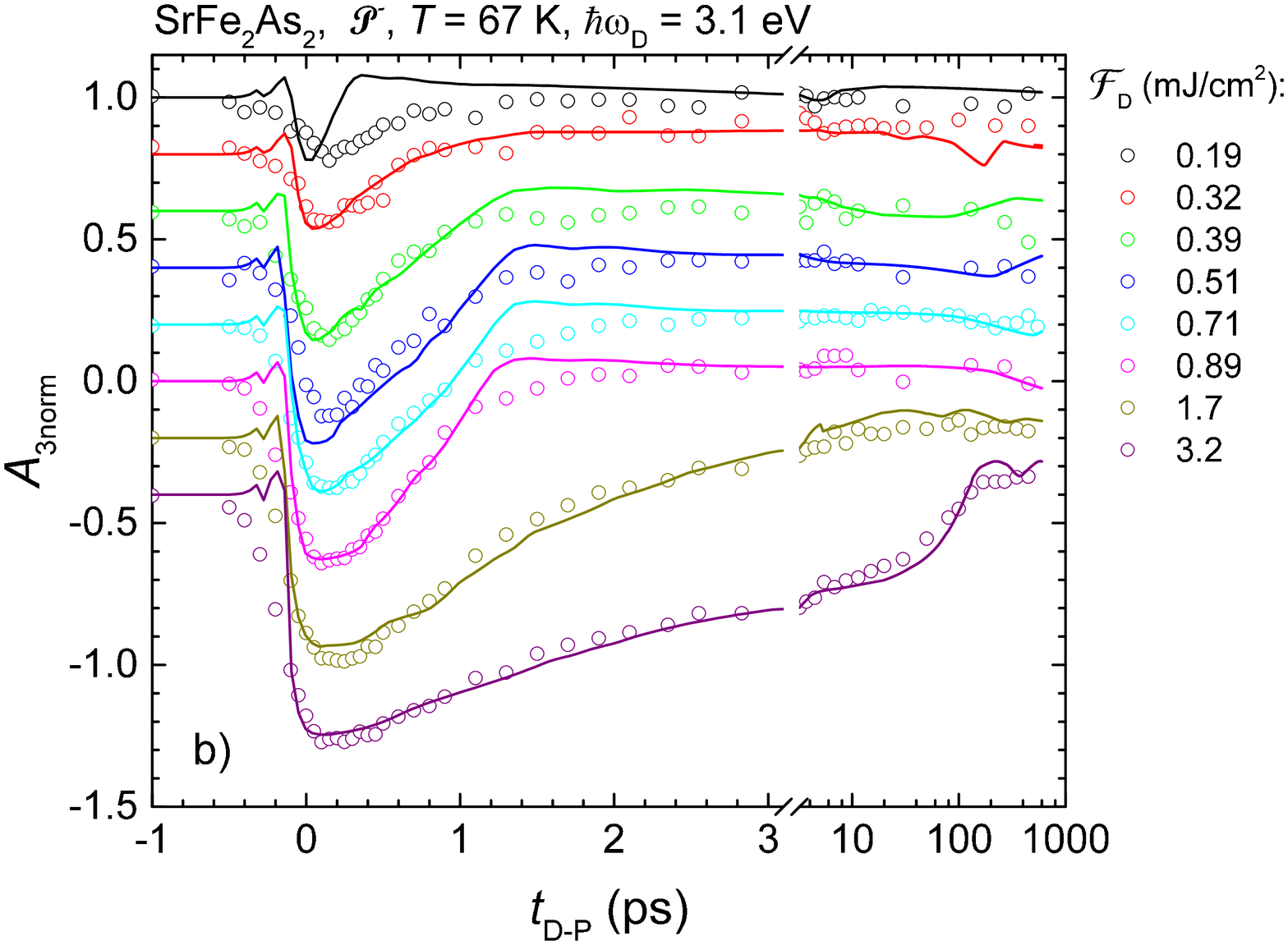}\caption{(Color online) Comparison of the experimental (open circles) and simulated
(lines) trajectories for the case of $\phi=\pi/2$. The trajectories
at different $\mathcal{F}_{\mathrm{D}}$ are vertically shifted for
clarity.}

\label{fig:fig-3T-sim-400} 
\end{figure*}

Since $A(T)$ is virtually temperature independent in the SDW state
dropping abruptly above $T_{\mathrm{SDW}}$ the trajectories, $A_{3}(t_{\mathrm{DP}}),$
are expected to reflect mainly the SDW volume-fraction dynamics in
the probed volume\footnote{Within the probe penetration depth of $\sim1/\alpha_{\mathrm{Pr}}\sim30$
nm.} and/or the normal/nematic-state dynamics when the SDW state is completely
suppressed. Since $A(T)\rightarrow0$ above $\sim300$ K the sensitivity
in the later case is limited to the temperature window between $\sim200-\sim300$
K.

The evolution of $T_{\mathrm{e}}(z,t_{\mathrm{DP}})$ after the D
pulse is calculated solving a three temperature model\cite{perfettiLoukakos2007,RettigCortes2013},
where a subset of optical phonons with temperature, $T_{\mathrm{o}}$,
different from the lattice temperature, $T_{\mathrm{L}}$, is assumed
to be strongly coupled to the electronic subsystem.\footnote{See Section \ref{sub:Three-temperature-model} for the detailed formulation.}
In the simplest 3TM it is also assumed that all the absorbed energy
remains in the electronic subsystem until it completely thermalizes.
Here the model is extended assuming that a part, $\eta$, of the absorbed
energy is transferred directly to a few strongly coupled optical phonons
during thermalization of the NEDF on a few-$100$-fs timescale.\cite{DemsarAveritt2003} 

To fit the simulated trajectories to the experimental data we take
the experimental specific heat capacity, $c_{\mathrm{p}}(T)$,\cite{Herero-MartinScagnoli2009,ChenLi2008},
and set $\phi$ in Eq. (\ref{eq:sat-dr-2D}) to the either 0 or $\pi/2$.
We also fix the heating pulse length to 200 fs corresponding to the
trajectory temporal resolution. The rest of the 3TM parameters are
determined from the nonlinear least squares fit. In the first step
only the trajectories for the highest $\mathcal{F}_{\mathrm{D}}$
are fit. Due to rather accurate total specific-heat-capacity\cite{Herero-MartinScagnoli2009,ChenLi2008}
and static optical-reflectivity data\cite{CharnukhaLarkin2013,WuBarisic2009}
the thermal conductivity, $k,$ and the phenomenological D-pulse saturated-absorption
length, $L_{\mathrm{D}}\sim80$ nm,\footnote{See Appendix, Eq. (\ref{eq:source}) for the formal definition.}
at the highest $\mathcal{F}_{\mathrm{D}}$ are obtained from the long
DP-delay behavior in both samples. 

To fit the lower-$\mathcal{F}_{\mathrm{D}}$ trajectories it is assumed
that $L_{\mathrm{D}}\propto\mathcal{F}_{\mathrm{D}}$\footnote{Setting $\mathcal{F}_{\mathrm{D}}\sim0$ in (\ref{eq:source}) leads
to approximately exponential fluence decay with the penetration depth
equal to the equilibrium optical penetration depth. } while a global fit over all trajectories at different $\mathcal{F}_{\mathrm{D}}$
is used to determine the remaining parameters. With all the fit parameters
taken to be independent of $\mathcal{F}_{\mathrm{D}}$ it is not possible
to obtain reasonable fits at all experimental $\mathcal{F}_{\mathrm{D}}$s
simultaneously since the 3TM model results in too strong slowdown
of the relaxation with increasing $\mathcal{F}_{\mathrm{D}}$. On
the other hand, assuming that $G_{\mathrm{ol}}$ and $\eta$ depend
on $\mathcal{F}_{\mathrm{D}}$ and setting $\phi=\pi/2$\footnote{Fits with $\phi=0$ show much worse agreement at low $\mathcal{F}_{\mathrm{D}}$. }
results in excellent fits (shown in Fig. \ref{fig:fig-3T-sim-400})
in the complete experimental $\mathcal{F}_{\mathrm{D}}$ range. The
quality of the fits supports the initial hypothesis of the fast sub
200-fs order parameter dynamics.

\begin{figure}
\includegraphics[width=0.65\columnwidth]{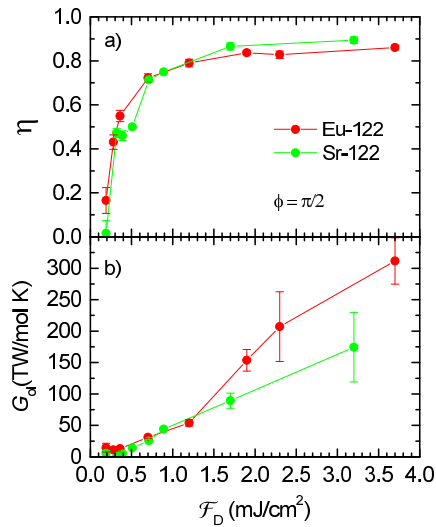}\caption{(Color online) The destruction-pulse fluence dependence of the branching
factor (a) and the electron - optical-phonons coupling (b) from the
3TM fit. \label{fig:3T-fit-par}}
\end{figure}

The obtained 3TM fit parameters are shown in Table \ref{tab:3Tpara}
and Fig. \ref{fig:3T-fit-par}. For comparison the fit parameters
from fits to the published time resolved -ARPES\cite{RettigCortes2013}
(TR-ARPES) surface-$T_{\mathrm{e}}$ dynamics in Eu-122 are also shown.

The obtained normal-state values of the electronic specific heat constant,
$\gamma_{\mathrm{e}}$, in the 50-60 mJ/mol~K$^{2}$ range (Table
\ref{tab:3Tpara}) are significantly larger than the low-temperature
(SDW-state) thermodynamic value of $\gamma_{\mathrm{e}}\sim8$~mJ/mol~K$^{2}$
\cite{ChenLi2008}.\footnote{We take $\gamma_{\mathrm{e}}$ measured in Sr-122 since in it is not
experimentally accessible Eu-122 due to the Eu$^{2+}$ spin ordering
at low $T$.} The increase of $\gamma_{\mathrm{e}}$ in the normal state is consistent
with the suppression of the SDW gap, but appears somewhat larger than
upon suppression of the SDW state by Co doping\cite{HardyBurger2010}
in Ba-122, where $\gamma_{\mathrm{e}}$ increases from $\sim5$ mJ/mol~K$^{2}$
in the SDW state to $\sim25$ mJ/mol~K$^{2}$ in the superconducting
samples. On the other hand, assuming that the high normal state magnetic
susceptibility\cite{Johnston2010} is dominated by the Pauli contribution
and the electron-phonon coupling constant is small\cite{StojchevskaKusar2010,StojchevskaMertelj2012}
results in comparable $\gamma_{\mathrm{e}}\sim60$~mJ/mol~K$^{2}$.

While the values of $\gamma_{\mathrm{e}}$ for Eu-122 obtained from
our data and TR-ARPES are consistent there is much larger discrepancy
of the other parameters. Fits to the multipulse trajectories result
in a smaller electron phonon relaxation rate, $G_{\mathrm{eo}}$,
larger optical-phonon - lattice relaxation rate, $G_{\mathrm{ol}}$,
and significantly smaller strongly-coupled optical-phonon heat capacity.
Partially this can be attributed to the systematic errors of the 3TM
and the response function. Setting $\eta$ to a fixed $\mathcal{F}_{\mathrm{D}}$-independent
value results in qualitatively similar trajectories (see Appendix
\ref{sub:3TM-fits-single}, Fig. \ref{fig:fig-3T-sim-400-single}
and Table. \ref{tab:3Tpara-single}) with similar $\gamma_{\mathrm{e}}$
and $c_{\mathrm{E0}}$, but significantly different relaxation rate
parameters. Another obvious contribution to the difference are also
differences between the surface and bulk since the present technique
is more bulk sensitive than TR-ARPES.

By using a simpler two temperature model with $\mathcal{F}_{\mathrm{D}}$-dependent
$\gamma_{\mathrm{e}}$ and the electron phonon coupling it is also
possible to obtain fair fits to the trajectories (not shown). However,
from such fits an nonphysically large $\gamma_{\mathrm{e}}$ of $\sim200$
~mJ/mol~K$^{2}$ is obtained indicating that some strongly coupled
optical phonons must play a role in the energy relaxation. It therefore
appears that the \emph{dominant relaxation bottleneck} is cooling
of the strongly-coupled optical phonons to the lattice bath. 

$\mathcal{F}_{\mathrm{D}}$-dependence of the optical-phonons - lattice
relaxation rate, $G_{\mathrm{ol}}$, shows a strong increase with
increasing $\mathcal{F}_{\mathrm{D}}$ (see Fig. \ref{fig:3T-fit-par}).
The increase is robust to the variations of the branching-factor fitting
approach (see Appendix \ref{sub:3TM-fits-single}, Fig. \ref{fig:3T-fit-par-single})
and can be attributed to opening of additional electronic relaxation
channels upon suppression of the nematic-fluctuations related pseudogap\cite{StojchevskaMertelj2012}
in addition to the anharmonic-decay channels.

A less robust\footnote{Possible excitation density dependence of the the response function
(\ref{eq:response-function}) could cause the worse fits using $\mathcal{F}_{\mathrm{D}}$-independent
$\eta$. } result of our analysis is the increase of the branching factor, $\eta$,
with increasing $\mathcal{F}_{\mathrm{D}}$ suggesting that above
$\mathcal{F}_{\mathrm{D}}\sim1$~mJ/cm$^{2}$ the majority of the
absorbed optical energy is on $\sim100$-fs timescale transferred
to the strongly coupled optical phonons. This is corroborated with
a quick initial recovery of the anisotropy (Fig. \ref{fig:fig-anis})
that indicates that $T_{\mathrm{e}}$ drops below $\sim300$ K into
the region of strong nematic fluctuations already a few hundred fs
after the arrival of the D-pulse. While the increase of $\eta$ appears
correlated with the observed optical nonlinearity we could not come
up with any persuasive physical picture to explain the effect so we
leave it open for further experimental confirmation and discussion.

\subsection{Destruction timescale}

The experimental destruction timescale of $\sim150$ fs could be set
either by the intrinsic low-energy SDW order parameter dynamics or
the finite initial NEDF thermalization timescale. While the intrinsic
SDW order dynamics on the $\sim150$-fs timescale would not contradict
the 3TM simulation results, the dependence of the destruction timescale
on the D-photon energy in Sr-122 suggests that the destruction timescale
is set by the initial NEDF thermalization.

\subsection{Determination of the SDW destruction threshold\label{sub:Determination-of-the}}

As in superconductors and charge density waves\cite{StojchevskaKusar2011}
we associate the saturation of the transient reflectivity amplitude
in the standard pump-probe experiments with destruction of the ordered
state. In the present case the saturation is incomplete where the
finite slope at high excitation density presumably corresponds to
the transient response of the normal, unordered state. 

The shape of the saturation curve {[}see Fig. \ref{fig:fig-F-dep}
(c) and (g){]} depends on the SDW destruction threshold excitation
energy density, $U{}_{\mathrm{th}}$, the geometrical parameters of
the pump and probe beams and their penetration depths.\footnote{The ordered state destruction is spatially non uniform due to the
inhomogeneous excitation profile. } In addition, the contribution of the pump-absorption saturation has
to be taken into account in the present case. 

To take into account the above effects we formulate a simple phenomenological
saturation model where we approximate the local amplitude of the transient
change of the dielectric constant, $\Delta\epsilon(r,z)$, by a piece-wise
linear function of the locally absorbed energy density, $U(r,z)$,
that has different slopes below and above $U{}_{\mathrm{th}}$:
\begin{eqnarray}
\Delta\epsilon(r,z) & = & \Delta\epsilon_{0}g(r,z),\nonumber \\
g(r,z) & = & \begin{cases}
\frac{U(r,z)}{U_{\mathrm{th}}}; & U(r,z)<U_{\mathrm{th}}\\
1+a(\frac{U(r,z)}{U_{\mathrm{th}}}-1); & U(r,z)\geq U_{\mathrm{th}}
\end{cases},\label{eq:sat}
\end{eqnarray}
where $r$ corresponds to the radial distance from the beam center,
$z$ the normal distance from the sample surface and $a$ to the relative
slope in the normal state. The spatial dependence of $U(r,z)$ is
given by: 
\begin{eqnarray}
\frac{U(r,z)}{U_{\mathrm{th}}} & = & \frac{\mathcal{F}_{0}}{\mathcal{F}_{\mathrm{th}}}\frac{\left(1+e^{-\alpha_{\mathrm{P}}L_{\mathrm{P}}(\mathcal{F}_{\mathrm{th}},0)}\right)e^{-2r^{2}/\rho_{\mathrm{P}}^{2}}}{\left(1+e^{\alpha_{\mathrm{P}}[z-L_{\mathrm{P}}(\mathcal{F}_{0},r)]}\right)},\\
L_{\mathrm{P}}(\mathcal{F},r) & = & c_{\alpha}\mathcal{F}e^{-2r^{2}/\rho_{\mathrm{P}}^{2}},\label{eq:lpvsr}
\end{eqnarray}
where $\alpha_{\mathrm{P}}$ is the linear pump absorption coefficient.
We phenomenologically take into account the pump-absorption saturation
by using the Fermi function to model the $U(r,z)$ depth dependence
introducing the local fluence dependent pump-penetration depth (\ref{eq:lpvsr}).
The coefficient, $c_{\alpha}$, is determined from the multi-pulse
experiment fits discussed above, while the pump beam is characterized
by the pump beam diameter, $\rho_{\mathrm{P}}$, and the external
fluence in the center of the beam, $\mathcal{F}_{0}$. $\mathcal{F}_{\mathrm{th}}$
corresponds to the external threshold fluence at which $U{}_{\mathrm{th}}$
is reached at the surface ($z=0$) in the center of the beam ($r=0$).

\begin{table}
\medskip{}
\begin{tabular}{c|c|c|c|c}
$\hbar\omega_{\mathrm{P}}$ & \multicolumn{2}{c|}{1.55 eV} & \multicolumn{2}{c}{3.1eV}\tabularnewline
\hline 
$\phi$ & 0 & $\pi/2$ & 0 & $\pi/2$\tabularnewline
\hline 
 & \multicolumn{4}{c}{$\mathcal{F}_{\mathrm{th}}$ (mJ/cm$^{2}$)}\tabularnewline
\hline 
EuFe$_{2}$As$_{2}$ & 0.21 & 0.15 & 0.18 & 0.12\tabularnewline
SrFe$_{2}$As$_{2}$ & 0.21 & 0.16 & 0.20 & 0.16\tabularnewline
\end{tabular}\caption{The external destruction threshold fluence, $\mathcal{F}_{\mathrm{th}}$
at different pump photon energies, $\hbar\omega_{\mathrm{P}}$ for
two extreme phase shifts $\phi$. The static optical constants used
in fits were taken from Refs. {[}\onlinecite{WuBarisic2009}{]} and
{[}\onlinecite{CharnukhaLarkin2013}{]}.}
\label{tab:Fth}
\end{table}

In the case of a relatively wide\footnote{With respect to the optical wavelength.}
Gaussian probe beam with diameter $\rho_{\mathrm{Pr}}$ Eq. (\ref{eq:sat-dr-2D})
describing the transient-reflectivity amplitude can be simply upgraded
to take into account the radial variation of the response (see also
Appendix \ref{sub:Transient-reflectivity}): \begin{widetext} 
\begin{eqnarray}
A & \propto\int_{0}^{\infty}rdr\int_{0}^{\infty}dz & e^{-2r^{2}/\rho_{\mathrm{Pr}}^{2}}e^{-\alpha_{\mathrm{Pr}}z}\cos\left(2n\frac{\omega_{0}}{c_{0}}z-\phi\right)g(r,z).\label{eq:sat-dr-3D}
\end{eqnarray}
\end{widetext} 

When fitting Eq. (\ref{eq:sat-dr-3D}) to the experimental data it
turns out that $\mathcal{F}_{\mathrm{th}}$, $\phi$ and $a$ are
strongly correlated. Since $\phi$ is usually not known \emph{a priori}
we fix $\phi$ to either 0 or $\pi/2$ to obtain a range of values
for $\mathcal{F}_{\mathrm{th}}$. Example fits with $\phi=\pi/2$,
are shown\footnote{Contrary to the multipulse trajectories simulations (see Fig. \ref{fig:fig-3T-sim-400}),
where taking $\phi=0$ resulted in worse fit to the data, the $\phi=0$
fit curves are virtually identical to the $\phi=\pi/2$ curves in
this case.} in Fig. \ref{fig:fig-F-dep} (c) and (g) with the resulting $\mathcal{F}_{\mathrm{th}}$
shown in Table \ref{tab:Fth}. While the variation of $\phi$ can
strongly influence the extracted $\mathcal{F}_{\mathrm{th}}$ the
determined ranges of $\mathcal{F}_{\mathrm{th}}$ are very similar
in both samples at both pump-photon energies.

Taking $\phi=\pi/2$ indicated by the 3TM simulations (Table \ref{tab:Fth})
we calculate the destruction treshold energy density, $U_{\mathrm{th}}\sim1.6$
kJ/mol for Eu-122 and $U_{\mathrm{th}}\sim2.5$ kJ/mol for Sr-122.
Assuming that $U_{\mathrm{th}}$ corresponds to the condensation energy
and taking $\gamma_{\mathrm{e}}$ from the 3TM fits we can estimate
the SDW gap using the standard BCS formula and obtain $\nicefrac{2\Delta_{\mathrm{SDW}}}{k_{\mathrm{B}}T_{\mathrm{SDW}}}=5$
and 6 for Eu-122 and Sr-122, respectively. This is somewhat lower
than the earlier weak-excitation pump-probe estimate\cite{PogrebnaVujicic2014}
of 13 and 8 for Eu-122 and Sr-122, respectively, but closer to the
optical conductivity result of 5.6 in Eu-122 \cite{WuBarisic2009}. 

Contrary to the superconductors\cite{StojchevskaKusar2011,BeyerStaedter2011}
there is no indication that the optical destruction energy would significantly
exceed the estimated SDW condensation energy. This is consistent with
the 3TM trajectories fit results where at small $\mathcal{F}_{D}$
that is comparable to $\mathcal{F}_{\mathrm{th}}$ the fast optical
energy transfer to the phonons is rather small (Fig. \ref{fig:3T-fit-par}).

\section{Summary and conclusions}

We presented an extensive all-optical study of the transient SDW state
suppression and recovery in EuFe$_{2}$As$_{2}$ and SrFe$_{2}$As$_{2}$
under strong ultrafast optical excitation by means of the standard
time resolved pump-probe as well as the multi-pulse transient optical
spectroscopy.

The SDW order is suppressed on a $\sim150$ - $\sim250$-fs timescale
after a $50$-fs destruction optical pulse absorption, depending on
the optical-photon energy. The suppression time scale is fluence independent
and set by the initial electronic thermalization timescale. 

The SDW recovery timescale increases with the destruction optical-pulse
fluence, but remains below $\sim1$ ps up to the fluence at which
the transient lattice temperature exceeds the SDW transition temperature. 

The optical SDW destruction threshold energy densities of $\sim1.6$
kJ/mol and $\sim2.5$ kJ/mol in EuFe$_{2}$As$_{2}$ and SrFe$_{2}$As$_{2}$,
respectively, are consistent with the BCS condensation energy estimates. 

The time evolution of the multi-pulse system trajectories in a broad
destruction-pulse fluence range can be well described within the framework
of an extended three temperature model assuming a fast sub 200-fs
intrinsic order parameter timescale. The model fits indicate the normal
state specific heat constant, $\gamma_{\mathrm{e}}$, in the 50-60
mJ/mol~K$^{2}$ range. The fluence-dependent recovery timescale is
found to be governed by the optical-phonons - lattice relaxation bottleneck
that is strongly suppressed at high excitation densities. The suppression
of the bottleneck is attributed to a suppression of the the nematic-fluctuations
induced pseudogap at high temperatures.

The observed resilience of the SDW state at high fluences exceeding
the SDW-destruction threshold fluence of $\sim0.15$ mJ/cm$^{2}$
up to $\sim10$ times is attributed to saturation of the optical absorption.
The model fits also suggest that at these fluences the majority of
the absorbed optical energy is transferred to the optical phonons
during the initial electronic thermalization on a few hundred femtosecond
time scale.
\begin{acknowledgments}
The authors acknowledge the financial support of Slovenian Research
Agency (research core funding No-P1-0040) and European Research Council
Advanced Grant TRAJECTORY (GA 320602) for financial support.
\end{acknowledgments}

\section{Appendix}

\subsection{Transient reflectivity\label{sub:Transient-reflectivity}}

Assuming that the beam diameters are large in comparison to the optical
penetration depth and the transient dielectric constant varies slowly
on the optical pulse timescale, $\Delta\epsilon(z,t)\sim\Delta\epsilon(z)$,
we can write the wave equation for the perturbed probe field in one
dimension:

\begin{equation}
\frac{\mathcal{N}^{2}}{c_{0}^{2}}\frac{\partial^{2}\Delta E}{\partial t^{2}}-\frac{\partial^{2}\Delta E}{\partial z^{2}}=-\frac{1}{\epsilon_{0}c_{0}^{2}}\frac{\partial^{2}\Delta P}{\partial t^{2}},\label{eq:wave}
\end{equation}
where $t$ is time, $z$ the distance from the sample surface, $\mathcal{N}=n+i\kappa$
the complex refraction index, $\Delta P$ the pump-probe induced transient
polarization with $c_{0}^{2}$ and $\epsilon_{0}$ the speed of light
and the vacuum permittivity, respectively. In the presence of a monochromatic
probe field, $E_{\mathrm{Pr}}(z,t)=t_{12}E_{\mathrm{0Pr}}e^{-i(\omega t-\mathcal{N}kz)}+\mathrm{c.c.}$,
propagating into the sample the transient polarization is given by:
\begin{eqnarray}
\Delta P(z) & = & \epsilon_{0}\Delta\epsilon(z)E_{\mathrm{Pr}}(z,t)=,\nonumber \\
 & = & t_{12}E_{\mathrm{0Pr}}\epsilon_{0}\Delta\epsilon(z)e^{-i(\omega t-\mathcal{N}\frac{\omega}{c_{0}}z)}+\mathrm{c.c.}\label{eq:transP}
\end{eqnarray}
where $E_{\mathrm{0Pr}}$ is the complex amplitude of the incident
probe field at the sample surface (at $z=0$) and $t_{12}=2/(1+\mathcal{N})$
is the Fresnel transmission coefficient. Solving (\ref{eq:wave})
assuming (\ref{eq:transP}) we obtain to the linear order in $\Delta\epsilon(z)$
the transient reflected field outside of the sample (at $z=0$): 
\begin{equation}
\Delta E_{\mathrm{r}}=\frac{i}{2}\frac{t_{12}t_{21}\omega}{c_{0}\mathcal{N}}E_{\mathrm{Pr0}}e^{-i\omega t}\int_{0}^{\infty}\Delta\epsilon(z)e^{2i\mathcal{N}\frac{\omega}{c_{0}}u}dz+\mathrm{c.c.},\label{eq:transE_CW}
\end{equation}
where the Fresnel coefficient $t_{21}=2\mathcal{N}/(1+\mathcal{N})$
takes into account the transmission from the sample to vacuum. 

In the case of an incident Gaussian probe pulse: 
\begin{eqnarray}
E_{\mathrm{Pr}}^{\mathrm{pulse}}(t) & =A_{0} & \sqrt{\frac{2}{\tau\sqrt{\pi}}}e^{-2t^{2}/\tau^{2}}e^{-i\omega_{0}t}+\mathrm{c.c.}=\nonumber \\
 & = & \int_{0}^{\infty}A(\omega-\omega_{0})e^{-i\omega t}d\omega+\mathrm{c.c.},
\end{eqnarray}
the total transient reflected electric field, neglecting the dispersion,
is the integral:\begin{widetext}

\begin{eqnarray}
\Delta E_{\mathrm{r}}^{\mathrm{pulse}}(t) & = & \frac{i}{2}\frac{t_{12}t_{21}}{c_{0}\mathcal{N}}\int_{0}^{\infty}dz\Delta\epsilon(z)\int_{0}^{\infty}d\omega\omega A(\omega-\omega_{0})e^{-i\omega t}e^{2i\mathcal{N}\frac{\omega}{c_{0}}z}+\mathrm{c.c.}=\nonumber \\
 & = & -\frac{t_{12}t_{21}A_{0}}{2\sqrt{2}\sqrt[4]{\pi}c_{0}\mathcal{N}\sqrt{\tau}}\frac{\partial}{\partial t'}\left[\int_{0}^{\infty}\Delta\epsilon(z)e^{-2t'^{2}/\tau^{2}}\times\right.\nonumber \\
 &  & \left.\times\left[e^{-i\omega_{0}t'}\left(1+\mathrm{erf}\left[\frac{\omega_{0}\tau}{2\sqrt{2}}+i\sqrt{2}\frac{t'}{\tau}\right]\right)+e^{i\omega_{0}t'}\left(1-\mathrm{erf}\left[\frac{\omega_{0}\tau}{2\sqrt{2}}-i\sqrt{2}\frac{t'}{\tau}\right]\right)\right]dz\right]+\mathrm{c.c.}\approx\nonumber \\
 & \approx & -\frac{t_{12}t_{21}A_{0}}{2\sqrt[4]{\pi}c_{0}\mathcal{N}\sqrt{\tau}}\frac{\partial}{\partial t}\left[\int_{0}^{\infty}\Delta\epsilon(z)e^{-2(t-2\mathcal{N}z/c_{0})^{2}/\tau^{2}}e^{-i\omega_{0}(t-2\mathcal{N}z/c_{0})}du\right]+\mathrm{c.c.},
\end{eqnarray}
where $t'=t-2\mathcal{N}z/c_{0}$. In the last line we assumed that
the pulse is narrowband, $\omega_{0}\tau\gg$1, so, 
\begin{equation}
\left|t'\right|/\tau\ll\omega_{0}\tau.\label{eq:timelimit}
\end{equation}
The transient reflectivity is then given by:
\begin{eqnarray}
\frac{\Delta R}{R} & = & \frac{\Delta I_{\mathrm{r}}}{I_{\mathrm{r}}}\simeq\frac{2\int\Re(\Delta E_{\mathrm{r}}^{\mathrm{pulse}}E_{\mathrm{r}}^{*})dt}{\int E_{\mathrm{r}}E_{\mathrm{r}}^{*}dt}=\nonumber \\
 & = & \Re\left[-\frac{t_{12}t_{21}}{\sqrt{\pi}c_{0}r_{12}\mathcal{N}\tau}\int dte^{-2t^{2}/\tau^{2}}e^{i\omega_{0}t}\times\right.\nonumber \\
 &  & \left.\times\frac{\partial}{\partial t}\int_{0}^{\infty}dz\Delta\epsilon(z)e^{i\omega_{0}(2\mathcal{N}z/c_{0}-t)}e^{-2(2\mathcal{N}z/c_{0}-t)^{2}/\tau^{2}}\right]=\nonumber \\
 & = & \Re\left[-\frac{\omega_{0}t_{12}t_{21}}{2c_{0}r_{12}\mathcal{N}}\int_{0}^{\infty}dz\Delta\epsilon(z)e^{2i\frac{\omega_{0}}{c_{0}}\mathcal{N}z}K(z)\right],\label{eq:drr}\\
K(z) & = & \left(1+i\frac{4\mathcal{N}z}{\omega_{0}c_{0}\tau^{2}}\right)e^{-4\mathcal{N}^{2}z^{2}/c_{0}^{2}\tau^{2}},\label{eq:K}
\end{eqnarray}
\end{widetext}where $E_{\mathrm{r}}=r_{12}E_{\mathrm{Pr}}^{\mathrm{pulse}}(t)$
and $r_{12}=(1-\mathcal{N})/(1+\mathcal{N})$ is the reflection coefficient.
In comparison to the CW case (\ref{eq:transE_CW}) an additional term
$K(z)$ appears in the kernel of the integral (\ref{eq:drr}). Due
to the exponent the kernel in (\ref{eq:drr}) decays on the length
scale $z\sim c_{0}/\omega_{0}\mathcal{\kappa}$ satisfying the condition
(\ref{eq:timelimit}) when $\omega_{0}\tau\gg\sqrt{n^{2}/\kappa^{2}+1}$.
On this length scale the argument of the exponent in (\ref{eq:K})
is of the order $4\left|\mathcal{N}\right|^{2}/\kappa^{2}\omega_{0}^{2}\tau^{2}\ll1$
for the narrowband pulses satisfying $\omega_{0}\tau\gg\sqrt{n^{2}/\kappa^{2}+1}$
so $K(z)\sim1$ can be dropped from (\ref{eq:drr}).

Due to the phase factor in (\ref{eq:drr}) the real and imaginary
parts of $\Delta\epsilon(z)=\Delta\epsilon_{\mathrm{r}}(z)+i\Delta\epsilon_{\mathrm{i}}(z)$
show different depth sensitivity that depends on the static complex
refraction index:\begin{widetext}

\begin{eqnarray}
\frac{\Delta R}{R} & = & \frac{4\omega_{0}}{c_{0}\left|\mathcal{N}^{2}-1\right|}\int_{0}^{\infty}dze^{-\alpha_{\mathrm{Pr}}z}\left[\Delta\epsilon_{\mathrm{r}}(z)\sin\left(2n\frac{\omega_{0}}{c_{0}}z-\beta\right)+\Delta\epsilon_{\mathrm{i}}(z)\cos\left(2n\frac{\omega_{0}}{c_{0}}z-\beta\right)\right],\label{eq:phaseFactors}\\
\tan(\beta) & = & \frac{\Im(\mathcal{N}^{2}-1)}{\Re(\mathcal{N}^{2}-1)}=\frac{2n\kappa}{n^{2}-\kappa^{2}-1},\nonumber 
\end{eqnarray}
\end{widetext}where, $\alpha_{\mathrm{Pr}}=2\kappa\frac{\omega_{0}}{c_{0}}$,
is the probe absorption coefficient.

\begin{figure}
\includegraphics[clip,width=1\columnwidth]{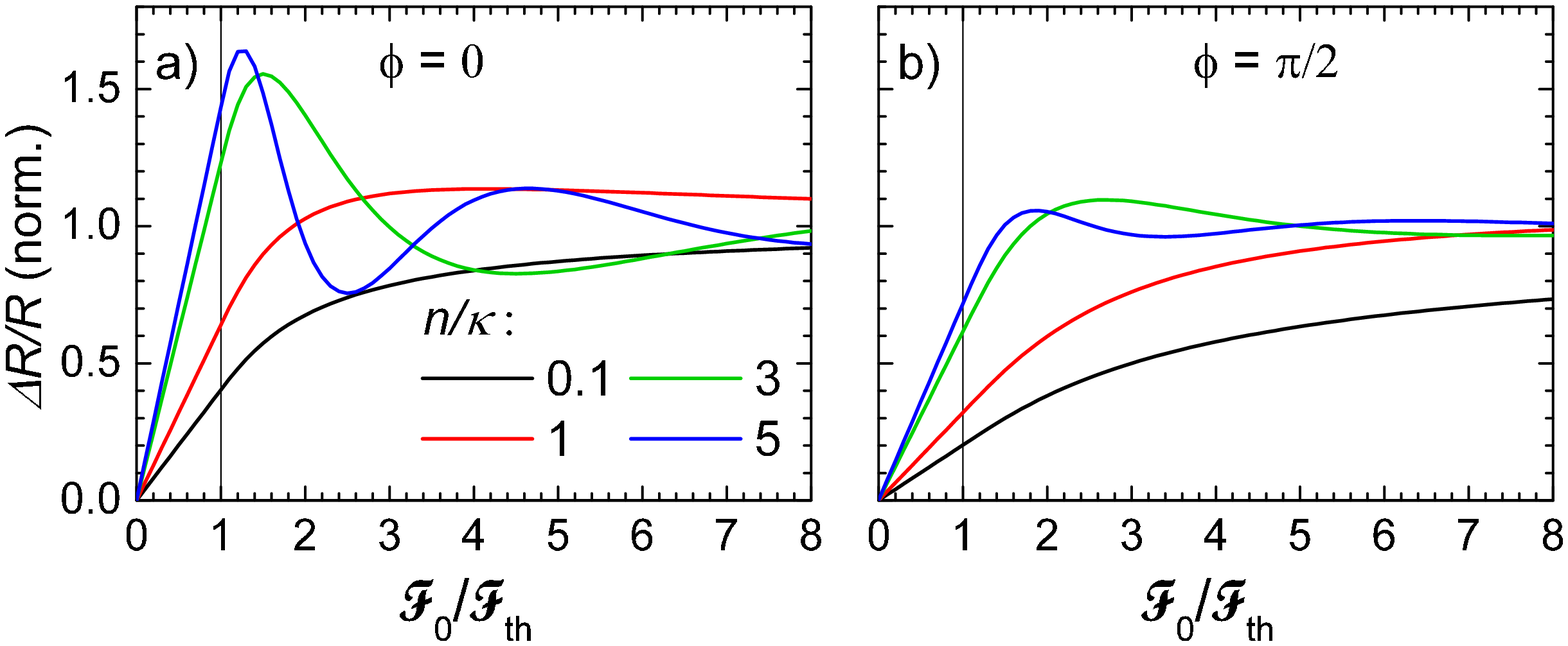}\caption{(Color online) Simulated fluence dependence for the case of Gaussian
beams from Eq. (\ref{eq:sat-dr-3D}) for different ratios $n/\kappa$
for two orthogonal phase shifts, $\phi$, taking $\rho_{\mathrm{P}}/\rho_{\mathrm{P}r}=2$,
$\alpha_{\mathrm{Pr}}/\alpha_{\mathrm{P}}=1$, $c_{\alpha}=0$ and
$a=0$. The curves are normalized to the saturated value at $\mathcal{F}_{0}\gg\mathcal{F}_{\mathrm{th}}$. }

\label{fig:fig-sat-sim} 
\end{figure}

Since in the simple saturation model (\ref{eq:sat}) the real and
imaginary parts of $\Delta\epsilon(z)$ are assumed to have the same
$z$ dependence Eq. (\ref{eq:phaseFactors}) is simplified to:\begin{widetext}
\begin{eqnarray}
\frac{\Delta R}{R} & = & \frac{4\omega_{0}\left|\Delta\epsilon_{0}\right|}{c_{0}\left|\mathcal{N}^{2}-1\right|}\int_{0}^{\infty}dze^{-\alpha_{\mathrm{Pr}}z}\cos\left(2n\frac{\omega_{0}}{c_{0}}z-\phi\right)g(z),\label{eq:sat-dr}\\
\tan(\phi) & = & \frac{2n\kappa\Delta\epsilon_{0\mathrm{r}}-(n^{2}-\kappa^{2}-1)\Delta\epsilon_{0\mathrm{i}}}{2n\kappa\Delta\epsilon_{0\mathrm{i}}+(n^{2}-\kappa^{2}-1)\Delta\epsilon_{0\mathrm{r}}}.\label{eq:phi}
\end{eqnarray}
\end{widetext}

In the case of Gaussian pump and probe beams with the diameters $\rho_{\mathrm{P}}$
and $\rho_{\mathrm{Pr}}$, respectively, (\ref{eq:sat-dr}) can be
easily extended\cite{KusarKabanov2008} to (\ref{eq:sat-dr-3D}) by
an additional integration in the radial direction\footnote{When both diameters are much larger than the corresponding wavelengths.}
where $g(r,z)$ is obtained from (\ref{eq:sat}) by taking into account
the radial pump fluence dependence. 

With increasing pump fluence the boundary between the ordered and
normal state region in (\ref{eq:sat}) moves along $z$, so the oscillatory
factor in the integral can lead to a non-monotonous excitation-density
dependence of the $\Delta R/R$ when $n/\kappa\gtrsim1$ as shown
in Fig. \ref{fig:fig-sat-sim}. There is unfortunately no clear singularity
observed when the threshold fluence is reached at low $n/\kappa$
ratios. Moreover, the saturation is much less pronounced for $\phi=\pi/2$.

\subsection{Three temperature model\label{sub:Three-temperature-model}}

\begin{figure}
\includegraphics[width=0.6\columnwidth]{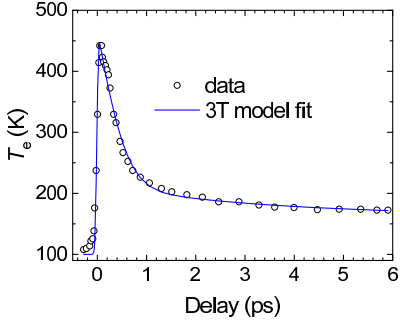}

\caption{Three temperature model fit to the electronic temperature in Eu-122
from TR-ARPES\cite{RettigCortes2013}\emph{. }Setting $\eta$ to different
values results in virtually identical fit curves (not shown) with
modified remaining 3TM parameters (see Table \ref{tab:3Tpara}).\emph{
\label{fig:TR-ARPES-fit}}}
\end{figure}

The time evolution of the temperatures in the three-temperature model
is governed by:
\begin{eqnarray}
\gamma_{\mathrm{e}}T_{\mathrm{e}}\frac{\partial T_{\mathrm{e}}}{\partial t} & = & (1-\eta)s(z,t)-G_{\mathrm{eo}}(T_{\mathrm{e}}-T_{\mathrm{o}})+\kappa\frac{\partial T_{\mathrm{e}}^{2}}{\partial z^{2}},\nonumber \\
c_{\mathrm{E}}(T_{\mathrm{o}})\frac{\partial T_{\mathrm{o}}}{\partial t} & = & \eta s(z,t)+G_{\mathrm{eo}}(T_{\mathrm{e}}-T_{\mathrm{o}})-G_{\mathrm{ol}}(T_{\mathrm{o}}-T_{\mathrm{l}}),\nonumber \\
c_{\mathrm{L}}(T_{\mathrm{L}})\frac{\partial T_{\mathrm{L}}}{\partial t} & = & G_{\mathrm{ol}}(T_{\mathrm{o}}-T_{\mathrm{L}}),
\end{eqnarray}
where $T_{\mathrm{e}}$, $T_{\mathrm{o}}$ and $T_{\mathrm{L}}$ are
the electronic, the strongly-coupled optical phonon (OP) and lattice
temperatures, respectively. $\gamma_{\mathrm{e}}$ is the normal state
electronic specific heat constant,\footnote{Since the response function (\ref{eq:response-function}) used for
calculating the trajectories is virtually constant below $T_{\mathrm{SDW}}$
we can neglect the drop of $\gamma_{\mathrm{e}}$ in the SDW state.} $c_{\mathrm{E}}(T_{\mathrm{o}})$ the Einstein phonon specific heat
and $c_{\mathrm{L}}(T_{\mathrm{L}})$ the lattice specific heat. $G_{\mathrm{eo}}$
and $G_{\mathrm{ol}}$ are the electron-OP and OP-lattice coupling
constants, while $\kappa$ is the electronic heat diffusivity. $s(z,t)$
is the absorbed laser energy density rate. Using the branching factor,
$\eta$, we take into account that the primary electron-hole pair
can relax by exciting the optical phonons during the thermalization.

To take into account absorption saturation we approximate $s(z,t)$
by:
\begin{equation}
s(z,t)\propto\mathcal{F}_{\mathrm{D}}e^{-\nicefrac{2t^{2}}{\tau_{\mathrm{p}}^{2}}}\left[1+e^{\alpha_{\mathrm{D}}(z-L_{\mathrm{D}})}\right]^{-1},\label{eq:source}
\end{equation}
where $\tau_{\mathrm{p}}$ is the effective heating pulse length,
$\alpha_{\mathrm{D}}$ the D-pulse linear absorption coefficient and
$L_{\mathrm{D}}$ the phenomenological absorption length.

$c_{\mathrm{E}}(T)$ is parametrized by the Einstein model:

\begin{equation}
c_{\mathrm{E}}(T)=c_{\mathrm{E}0}\left(\frac{T_{\mathrm{E}}}{T}\right)^{2}e^{\frac{T_{\mathrm{E}}}{T}}/(e^{\frac{T_{\mathrm{E}}}{T}}-1)^{2},
\end{equation}
while $c_{\mathrm{L}}(T)$ is obtained from the total experimental
specific heat capacity\cite{Herero-MartinScagnoli2009,ChenLi2008},
$c_{\mathrm{p}}(T)$, by subtracting the electronic and OP parts:
\begin{equation}
c_{\mathrm{L}}(T)=c_{\mathrm{p}}(T)-\gamma_{\mathrm{e}}T_{\mathrm{e}}-c_{\mathrm{E}}(T).
\end{equation}

According to Allen\cite{allen1987theory} the second moment of the
Eliashberg function can be expressed as:

\[
\lambda\left\langle \omega^{2}\right\rangle =\frac{\pi k_{\mathrm{B}}}{3\hbar}\frac{G_{\mathrm{eo}}}{\gamma_{\mathrm{e}}}.
\]

\subsection{3TM fits with a single $\mathcal{F}_{\mathrm{D}}$-dependent parameter\label{sub:3TM-fits-single}}

\begin{figure*}[tbh]
\includegraphics[clip,width=1\columnwidth]{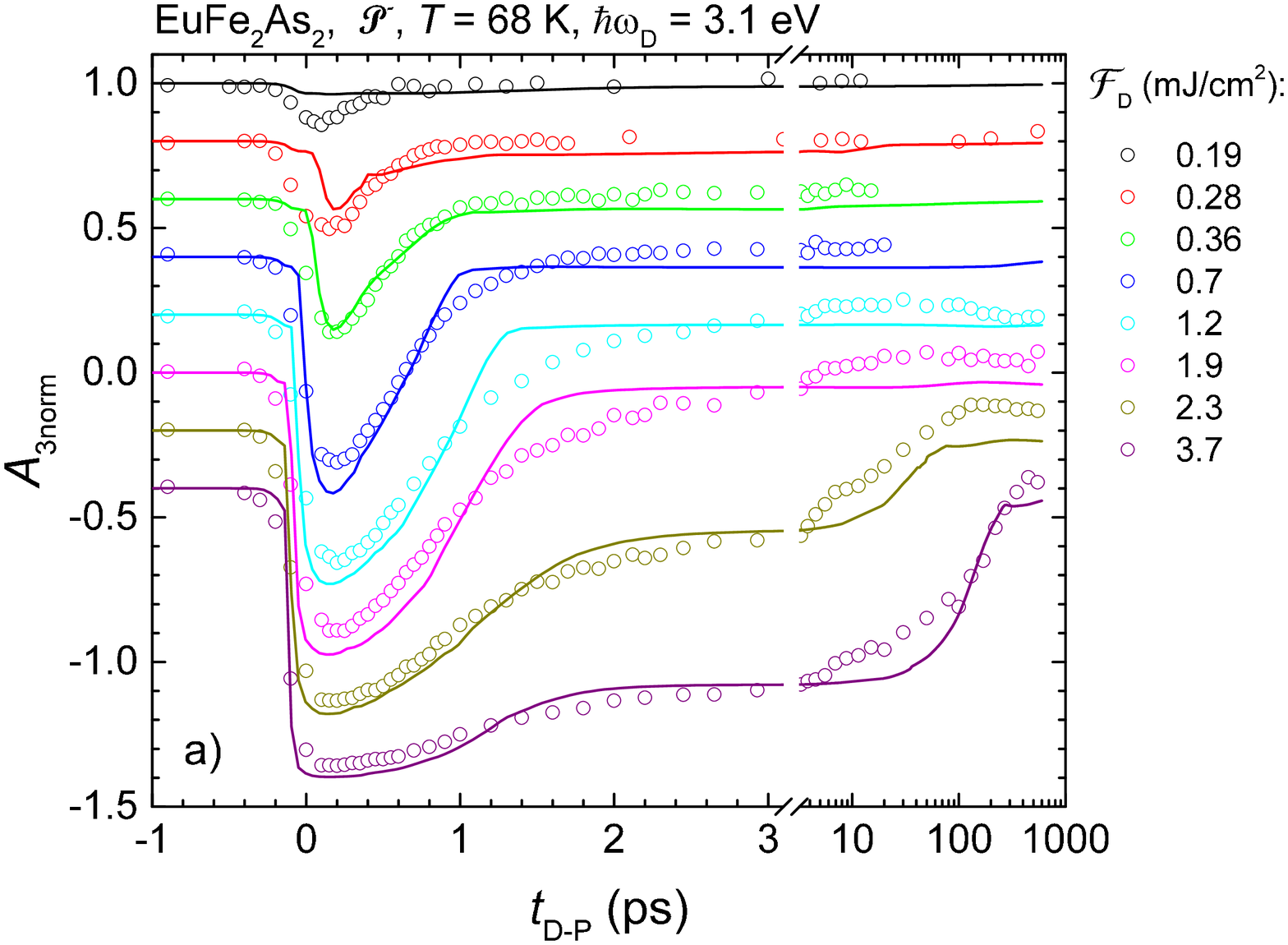}\includegraphics[clip,width=1\columnwidth]{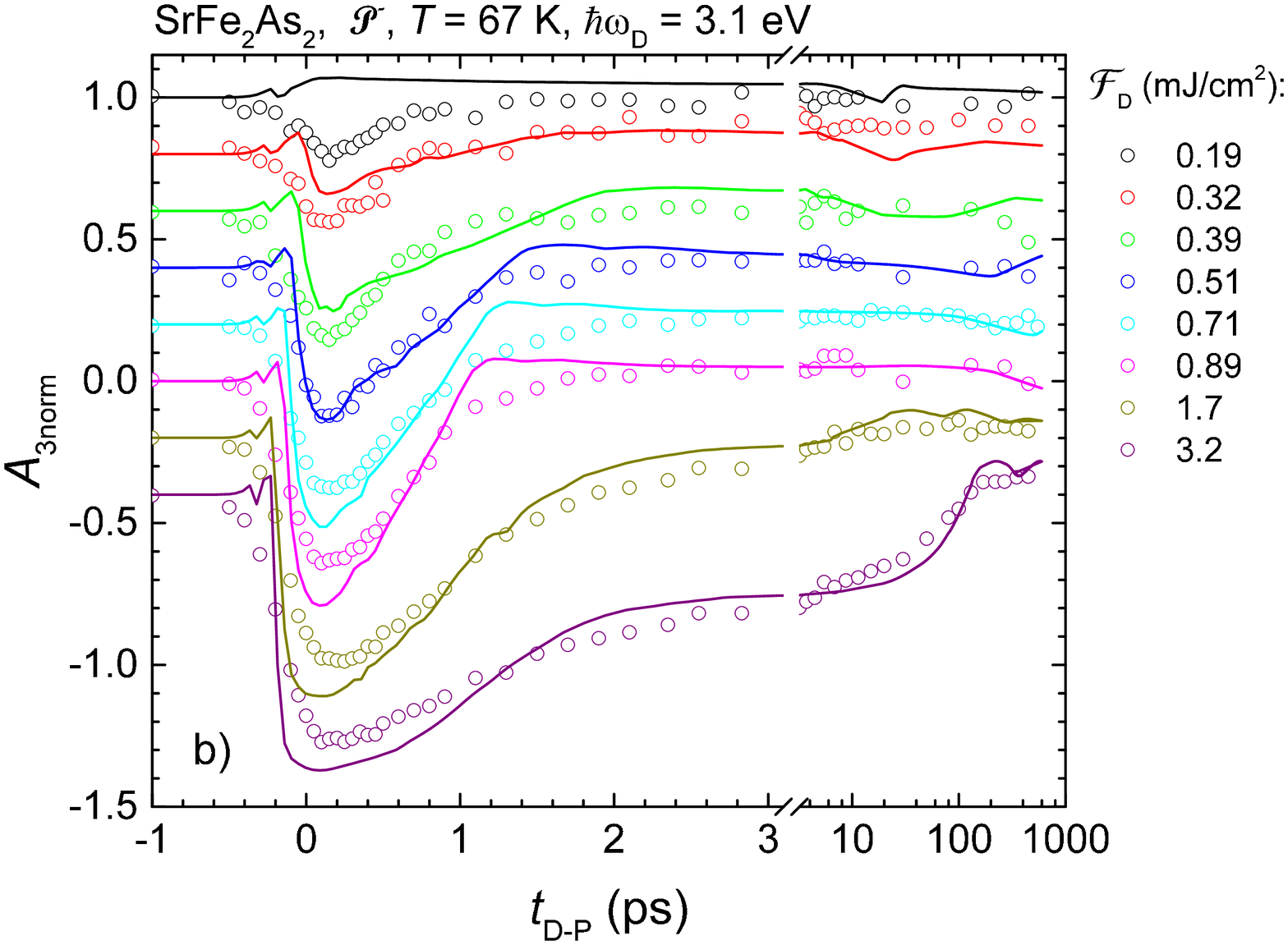}\caption{(Color online) Comparison of the experimental and simulated trajectories
for the case of $\eta=0.01$ and $\phi=\pi/2$. The trajectories at
different $\mathcal{F}_{\mathrm{D}}$ are vertically shifted for clarity.}

\label{fig:fig-3T-sim-400-single} 
\end{figure*}

\begin{figure}
\includegraphics[width=0.6\columnwidth]{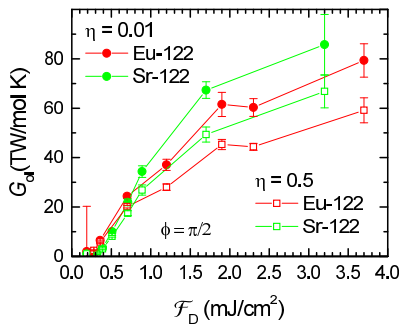}\caption{(Color online) The destruction-pulse fluence dependence of the electron
- optical-phonons coupling from the 3T model fit for two different
fixed branching factors. \label{fig:3T-fit-par-single}}
\end{figure}

A worse fit, particularly at low $\mathcal{F}_{\mathrm{D}}$, with
fixed and $\mathcal{F}_{\mathrm{D}}$-independent $\eta$ shown in
Fig. \ref{fig:fig-3T-sim-400-single} results in parameters shown
in Table \ref{tab:3Tpara-single}. The increase of $G_{\mathrm{ol}}$
with $\mathcal{F}_{\mathrm{D}}$ appears to be robust with respect
to the way how $\eta$ is fit (Fig. \ref{fig:3T-fit-par-single}).

\begin{table*}
\begin{tabular}{>{\raggedright}m{3.7cm}|c|c|c|c|c|c|c|c}
 & $\eta$\footnote{Fixed at the selected values.} & $\gamma_{\mathrm{e}}$ & $G_{\mathrm{eo}}$  & $\lambda\left\langle \omega^{2}\right\rangle $ & $c_{\mathrm{E}0}$\footnote{$T_{\mathrm{E}}$ was without fitting set to 300K.}  & $G_{\mathrm{ol}}$ & $k=\kappa/V_{\mathrm{mol}}$  & $L_{\mathrm{D}}/\mathcal{F}_{\mathrm{D}}$\footnote{Obtained from the two highest fluences fit in the multipulse case.}\tabularnewline
 & - & mJ/mol K$^{2}$  & TW/mol K  & (meV)$^{2}$ & J/mol K  & TW/mol K  & W/m K & nm cm$^{2}$/mJ\tabularnewline
\hline 
\multirow{2}{3.7cm}{Eu-122 (TR-ARPES)\cite{RettigCortes2013}\\
($T\sim100$ K) \footnote{ $\mathcal{F}_{\mathrm{D}}\sim1$ mJ/cm$^{2}$}} & $0.01$ & $52\pm3$ & $34\pm3$ & $39\pm3$ & $67\pm14$ & $7\pm5$ & - & 17\tabularnewline
\cline{2-9} 
 & 0.5 & $30\pm2$ & $23\pm3$ & $45\pm3$ & $76\pm20$ & $11\pm9$ & - & 17\tabularnewline
\hline 
\multirow{2}{3.7cm}{Eu-122 (present work)\\
($T\sim70$ K) } & 0.01 & $42\pm2$ & $96\pm6$ & $135\pm7$ & $15\pm1$ & \multirow{2}{*}{Fig. \ref{fig:3T-fit-par-single} } & \multirow{2}{*}{$14\pm1$} & \multirow{2}{*}{$17\pm5$}\tabularnewline
\cline{2-6} 
 & 0.5 & $10\pm1$ & $138\pm6$ & $820\pm130$ & $21\pm1$ &  &  & \tabularnewline
\hline 
\multirow{2}{3.7cm}{Sr-122 (present work)\\
($T\sim70$ K) } & 0.01 & $68\pm4$ & $112\pm6$ & $97\pm6$ & $17\pm1$ & \multirow{2}{*}{Fig. \ref{fig:3T-fit-par-single}} & \multirow{2}{*}{$10\pm1$} & \multirow{2}{*}{$23\pm5$}\tabularnewline
\cline{2-6} 
 & 0.5 & $38\pm1$ & $218\pm16$ & $340\mbox{\ensuremath{\pm}20}$ & $25\pm1$ &  &  & \tabularnewline
\end{tabular}\caption{Comparison of the three temperature model parameters obtained from
fits with $\mathcal{F}_{\mathrm{D}}$-dependent $G_{\mathrm{ol}}$.
\label{tab:3Tpara-single}}
\end{table*}

\bibliography{biblio}

\end{document}